%% file: msrRT17.tex
\documentclass[10pt, conference]{IEEEtran}
\ifCLASSINFOpdf
\else
\fi
\let\labelindent\relax
\usepackage[utf8]{inputenc}
\usepackage{booktabs}
\usepackage{amsmath}
\usepackage{amssymb}
\usepackage{paralist}
\usepackage{multirow}
\usepackage{xspace}
\usepackage{color}
\usepackage{xcolor}
\usepackage{ifthen}
\usepackage{url}
\usepackage{fancybox}
\usepackage{enumitem}
\usepackage{listings}
\usepackage{balance}
\usepackage{ifthen}
\usepackage[tight,footnotesize]{subfigure}
\usepackage{graphicx}
\usepackage{amsmath}
\usepackage[linesnumbered]{algorithm2e}
\usepackage{tablefootnote}
\input{macros}
\usepackage{flushend}

\graphicspath{{pics/}}

\definecolor{codegreen}{rgb}{0,0.6,0}
\definecolor{codegray}{rgb}{0.5,0.5,0.5}
\definecolor{codepurple}{rgb}{0.58,0,0.82}
\definecolor{backcolour}{rgb}{0.95,0.95,0.92}

\lstdefinestyle{mystyle}{
	backgroundcolor=\color{backcolour},   
	commentstyle=\color{codegreen},
	keywordstyle=\color{magenta},
	numberstyle=\tiny\color{codegray},
	stringstyle=\color{codepurple},
	basicstyle=\footnotesize,
	breakatwhitespace=false,         
	breaklines=true,                 
	captionpos=b,                    
	keepspaces=true,                 
	numbers=left,                    
	numbersep=2pt,                  
	showspaces=false,                
	showstringspaces=false,
	showtabs=false,                  
	tabsize=2
}

\usepackage[english]{babel}

\DeclareGraphicsExtensions{.pdf,.jpeg,.png}
\begin{document}

\newtheorem{theorem}{Definition}[section]
	
	\newcommand{\eg}{e.g.,}
	\newcommand{\ie}{i.e.,}

	\newcommand{\one}{small library}
	
	\newcommand{\ones}{small libraries}
	
	\newcommand{\tool}{DepChainsJS}
	\newcommand{\twoandtwoplus}{2nd and 3rd}

	\newcommand{\RqOne}{\emph{What is the size of a \texttt{npm} package?}\xspace}
	
	\newcommand{\RqTwo}{\emph{What is the dependency chain size for a \texttt{npm} package?}\xspace}
	
	\newcommand{\RqThree}{\emph{What is the developer usage cost of a \texttt{npm} package?}\xspace}
	
	\renewcommand{\lstlistingname}{Listing}


\title{On the Impact of Micro-Packages: An Empirical Study of the \texttt{npm} JavaScript Ecosystem}

\author{
	\begin{tabular}{ccc}
		\multicolumn{2}{c}{Raula Gaikovina Kula$^\ast$,  Ali Ouni$^\ast$, Daniel M. German$^\ddagger$ and  Katsuro Inoue$^\ast$
			\vspace{0.3cm}} \\
		{$^\ast$Osaka University, Japan} & {$^\ddagger$ University of Victoria, Canada}\\
		{\{raula-k, ali, inoue\}@ist.osaka-u.ac.jp} & {dmg@turingmachine.org} \\  
	\end{tabular} 
}


%

\maketitle

\begin{abstract}
The rise of user-contributed Open Source Software (OSS) ecosystems demonstrate their prevalence in the software engineering discipline.
Libraries work together by depending on each other across the ecosystem. 
From these ecosystems emerges a minimized library called a micro-package. 
Micro-packages become problematic when breaks in a critical ecosystem dependency ripples its effects to unsuspecting users. 
In this paper, we investigate the impact of micro-packages in the \texttt{npm} JavaScript ecosystem. 
Specifically, we conducted an empirical investigation with 169,964 JavaScript \texttt{npm} packages to understand (i) the widespread phenomena of micro-packages, (ii) the size dependencies inherited by a micro-package and (iii) the developer usage cost (ie., fetch, install, load times) of using a micro-package. 
Results of the study find that micro-packages form a significant portion of the \texttt{npm} ecosystem. 
Apart from the ease of readability and comprehension, we show that some micro-packages have long dependency chains and incur just as much usage costs as other npm packages. 
We envision that this work motivates the need for developers to be aware of how sensitive their third-party dependencies are to critical changes in the software ecosystem. 
\end{abstract}


%
\IEEEpeerreviewmaketitle

\section{Introduction}
\label{sec:intro}
User-contributed Open Source Software (OSS) ecosystems have become prevalent in the software engineering discipline, capturing the attention of both practitioners and researchers alike. 
In recent times, \textit{‘collections of third-party software’} ecosystems such as the node package manager \texttt{(npm)} JavaScript \cite{NPM} package ecosystem fosters development for huge amounts of server-side NodeJs and client-side JavaScript applications.
According to a study in 2016 \cite{Wittern:2016}, the \texttt{npm} ecosystem for the NodeJs platform hosts over 230 thousand packages with \textit{`hundreds of millions package installations every week}’. 
 
 
Jansen \textit{et al.} \cite{Jansen09ICSE} states that ecosystems emerge when \textit{`large computation tasks [are] split up and shared by collection of small, nearly independent, specialized units depending on each other'}. 
Based on this concept, we conjecture that third-party software ecosystems like \texttt{npm} encourage the philosophy of specialized software within these self-organizing ecosystems.
A \textit{micro-package} is the result of when a package becomes \textit{‘minimalist’} in its size and performs a single task \cite{Mens:2016}. 
For instance, the \texttt{negative-zero}\footnote{\url{https://www.npmjs.com/package/negative-zero}} package has the trivial task of determining whether or not an input number has a negative-zero value. 
Micro-packages function as a single unit by forming \textit{`transitive'} dependencies between dependent packages (\ie~dependency chains) across the ecosystem. 

We conjecture that an influx of micro-packages will result in an ecosystem that becomes fragile to any critical dependency changes.
This is where breaking one critical dependency in the ecosystem will ripple its effect down the dependency chain to all dependent packages.
For example, a breakage by removal of a tiny package called \texttt{left-pad} in March 2016 caused waves of dependency breakages throughout the \texttt{npm} ecosystem when it \textit{`broke the internet'} of web applications \cite{npmIssue}.
Although \texttt{left-pad} is tasked with simply \textit{‘adding left space padding to a html page’}, it was heavily relied upon in the ecosystem by thousands of unaware packages and applications, including the core \texttt{Babel} compiler and \texttt{Node} JavaScript environment.

In this paper, we investigate the impact of micro-packages within the \texttt{npm} JavaScript ecosystem. 
Specifically, we conducted an empirical investigation with 169,964 \texttt{npm} packages to understand (\textit{RQ1:}) the spread of micro-packages across the \texttt{npm} ecosystem. 
We then investigate micro-package usage implications such as the (\textit{RQ2:}) the size of its complex dependency chains and (\textit{RQ3:}) the amount of developer usage costs (ie., fetch, install, load times) incurred by using a micro-package.
Findings show that micro-packages account for a significant portion of the ecosystem, with some micro-packages having just as long dependency chain lengths and reachability to other packages in the \texttt{npm} ecosystem.
Furthermore, our findings indicate that micro-packages have no statistical differences in developer usage costs to the rest of the \texttt{npm} packages.
We envision that this work motivates the notion of micro-packages and how to increase their  resilience to changes in the ecosystem. 
The study concludes that developers should be aware of how sensitive their third-party dependencies are to critical changes in its ecosystem.

The main contributions of this paper are three-fold and can be summarized as follows: (i) we quantitatively study the phenomena of micro-packages and their impact in a software ecosystem, (ii) we motivate the need to revisit best practices to increase software resilience to their ecosystem changes and  (iii) we motivate the need for developers to be aware of how sensitive their third-party dependencies are to changes in a software ecosystem. 

\section{Basic Concepts \& Definitions}
\label{sec:background}
In this section, we introduce our definition of the different types of \texttt{npm} micro-packages and motivate how it can be problematic for developers that use these packages. 
We then introduce and define two types of micro-packages that we believe create excessive dependencies.

\subsection{Micro-Packages and a Fragile Ecosystem}
We define micro-packages as having a minimalist modular design through encapsulating and delegating complex tasks to other packages within the same ecosystem.
Encapsulating complexity through abstraction or information hiding to ease code comprehension and promotes code reuse.
Parnas \cite{Parnas:2002} first proposed concepts of information hiding through the formation of modular structures. 
Baldwin and Clark later \cite{Baldwin:1999} propose modularity theory, when elements of a design become split up and assigned to modules according to a formal architecture or plan.
In this setting, \textit{`some of these modules remain hidden, while other modules are visible and embody design rules that hidden modules must obey if they are to work together'.}

We conjecture that an influx of micro-packages will result in a fragile ecosystem. 
As packages evolve, so do their dependencies within this ecosystem. 
Sometimes these changing dependencies breaks a critical dependency that will impact and ripple changes throughout the ecosystem. 
Mens \textit{et al.} \cite{Mens:2016} reports experiences from both practitioners and researchers where managing the complexities of dependencies (i.e., colloquially known as \textit{Dependency Hell}) leads to issues that ripple throughout the ecosystem. 
Our hypothesis is that a micro-package contributes to a fragile ecosystem by creating excessive dependencies across the ecosystem.

\subsection{Package Interoperability in the \texttt{npm} Ecosystem}
The JavaScript language adheres to certain specifications\footnote{CommonJS Group has with a goal of building up the JavaScript ecosystem with specifications. Documentation at \url{http://wiki.commonjs.org/wiki/Modules/1.0}} that state how a package  should be written in order to be \textit{`interoperable among a class of module systems that can be both client and server side, secure or insecure, implemented today or supported by future systems with syntax extensions'.}

For the \textit{npm} ecosystem, according the \texttt{npmjs}\footnote{Official Website of the npmjs repository at \url{https://docs.npmjs.com/how-npm-works/packages}} documentation and \texttt{CommonsJS} Group\footnote{\url{http://wiki.commonjs.org/wiki/Packages/1.0}} specify that a \texttt{npm} package should at least have the following requirements:
\begin{itemize}
	\item a \texttt{package.json} configuration file.
	\item a JavaScript {module}.
	\item an accessible url that contains a gzipped tarball of \texttt{package.json} and the module.
\end{itemize} 
A \texttt{npm} module refers to any JavaScript file that can be loaded using the \texttt{require()} command.
Typically, the \texttt{package.json} points to an \texttt{entry point file}.
This entry point file is used to load any modules (\ie~both local and foreign) required by the package.

Importantly, the JavaScript \texttt{require()} function is unable to differentiate between a module or package.
Therefore, packages can load internal or external modules/packages that are installed on the \texttt{NodeJs} platform environment.
For external dependencies, the \texttt{require()} function enables interoperability between all declared global functions to be accessible as the Application Programming Interface (API).

\subsection{Micro-packages in the \texttt{npm} ecosystem}
We now introduce two types of \texttt{npm} micro-packages.
We first define two types of \texttt{npm} micro-packages that (i) perform trivial tasks and (ii) acts as a facade to load foreign module (\ie~third-party) dependencies.

\begin{lstlisting}[language=Python,
caption={Entry point source code of a the \texttt{negative-zero} library to \texttt{`Check if a number is negative zero'}.},
label=code:useful]
	'use strict';
	module.exports = function (x) {
		return x === 0 && 1 / x === -Infinity;
};	\end{lstlisting}

Listing \ref{code:useful} shows an example of a micro-package that performs a trivial task. 
In only four lines of code, the \texttt{negative-zero} package has the sole task of determining whether an input number has a value of negative zero. 
Importantly, the definition of a negative-zero is not prone to change so it is safe to assume that this package is less likely to be evolved in the near future.

\begin{lstlisting}[language=Python,
caption={The \texttt{user-home} package entry point source code. As shown the library depends on the \texttt{os-homedir} library.},
label=code:userhome]
	'use strict';
	module.exports = require('os-homedir')();\end{lstlisting}

Listing \ref{code:userhome} is example of a micro-package that forms a facade to another foreign package.
As shown in the listing, \texttt{user-home} package is comprised of only two lines of code, but is tasked with determining the user-home folder for any operating system. 
Furthermore, the listing shows (Line 2) the \texttt{user-home} library dependent and loading the \texttt{os-homedir} package (shown in Listing \ref{code:os-homedir}) to perform  complexities of the task.

\begin{lstlisting}[language=Python,
caption={\texttt{os-homedir} package entry point source code. It contains one function \texttt{homedir}.},
label=code:os-homedir]
	'use strict';
	var os = require('os');
	
	function homedir() {
	var env = process.env;
	var home = env.HOME;
	var user = env.LOGNAME || env.USER || env.LNAME || env.USERNAME;
		
	if (process.platform === 'win32') {
		return env.USERPROFILE || env.HOMEDRIVE + env.HOMEPATH || home || null;
	}
		
	if (process.platform === 'darwin') {
		return home || (user ? '/Users/' + user : null);
	}
	
	if (process.platform === 'linux') {
		return home || (process.getuid() === 0 ? '/root' : (user ? '/home/' + user : null));
	}
		
		return home || null;
	}
	module.exports = typeof os.homedir === 'function' ? os.homedir : homedir;\end{lstlisting}


Listing \ref{code:os-homedir} shows hidden \texttt{os-homedir} package that performs the computation for the \texttt{user-home} package. 
Importantly, the \texttt{user-home} package is prone to evolutionary dependency changes as operating systems (see Lines 9, 13, 17) may change their platform configurations in the future. 
In this case, any application that uses package with a dependency on the \texttt{user-home} is unaware that it is indirectly impacted by any evolutionary API changes or removal of the \texttt{user-home} package.

We leverage the package size attributes of (i) global function APIs and (ii) physical size to propose a method to identify the two types of micro-package defined in this study.
In detail, we propose and rationalize these package size metrics:

\begin{itemize}
	\item \textit{\# of FuncCount (FuncCount)} - is a count of API functions implemented in a package (\ie~function in the \texttt{entry point file}).
	Our rationale is a single function is more likely a trivial micro-package, while a facade micro-package is likely to use the \texttt{require()} function to directly load its dependencies.
	\item \textit{\# of Lines of Code (SLoC)} - is the count of source code lines in a package (\ie~function in the \texttt{entry point file}).
	We	consider both the physical (i.e., physical includes code comments and spacing) and logical lines of code for a	library. Our rationale is that trivial or facade micro-packages tend to have smaller sLoC.
\end{itemize}

We can now use these metrics to rationalize our definition of a micro-package. 
Formally, for any given Package $P$, $P$ is a micro-package when \texttt{funcCount(P)}$\le $ 1.


\section{Empirical Study}
\label{sec:study}
Our main goal is to better understand the role and impact of micro-packages in the \texttt{npm} ecosystem. 
We conducted an empirical investigation with three guiding research questions regarding micro-packages.

RQ1 relates to how widespread the phenomena of micro-packages exists in the \texttt{npm} ecosystem. 
In this research question analyzes the size of \texttt{npm} packages to determine whether micro-packages form a significant portion of the ecosystem.

RQ2 and RQ3 relates the evaluation of some perceptions on when using a micro-package.
From a complexity of dependencies viewpoint, RQ2 investigate the size of dependency chains developers incur when using a micro-package. 
For RQ3, we then examine fetch, install and load usage costs incurred by a web developer when using a micro-package as one of its dependencies.

\subsection{\textbf{RQ1: \RqOne}}
\subsubsection{\underline{Motivation}}
Our motivation for RQ1 is to understand how widespread the micro-package phenomena is in the ecosystem. 
In this research question, we analyze \texttt{npm} package size attributes to survey whether or not micro-packages form a significant portion of packages in the ecosystem.

\subsubsection{\underline{Research Method}}
Our research method to answer RQ1 is by statistical analysis and manual validation. 
It follows a two step of data collection and then analysis of the results.
In the first step, we collect and process \texttt{npm} libraries to extract their size attributes (\ie~FuncCount and SLoC) for each package.

For the analysis step, we describe the npm size metrics distribution using statistical metrics (\ie~Min., Mean ($\mu$), Median ($\bar{x}$), Max.) and plot on a histogram. 
We then manually investigate and pull out examples of the results to qualitatively explain the reasoning behind our conclusions.
Finally, we identify how many npm packages fit our definition of micro-packages.
We then utilize boxplots plots with the sLoC metric to compare any statistical differences between micro-packages to the rest of the packages. 




\label{sec:rm1}

\label{sec:data1}
\begin{table}
	\begin{center}
		\caption{Dataset Collected \texttt{npm} corpus for RQ1
		}
		\label{tab:datasetRQ1}
		\begin{tabular}{lrr}
			\hline
			\multirow{1}{*}{} 
			& \multirow{1}{*}{\textbf{Dataset statistics}}
			\\
			Snapshot of \texttt{npm} ecosystem&July-1st-2016 $\sim$ July-15th-2016&\\
			\# downloads &186,507 libraries \\
			Size (GB)&200GB\\  
			\# js entry point files analyzed& 169,964 files\\\hline
		\end{tabular}
	\end{center}
\end{table}

\subsubsection{\underline{Dataset}}
Similar to the collection method of \cite{Wittern:2016}, we use the npm registry\footnote{\url{https://registry.npmjs.org/-/all}} to procure an offline copy of the publicly available packages \textit{npm} ecosystem. 

Table \ref{tab:datasetRQ1} presents statistics of the final collected dataset after a quality pre-processing was performed to remove invalid or erroneous \texttt{package.json} metafiles\footnote{A blog on the official npmjs website details some of the perils of the npm front-end packaging at \url{http://blog.npmjs.org/post/101775448305/npm-and-front-end-packaging}}.
As shown in the table, after downloading 186,507 \texttt{npm} packages, 169,964 packages remain for our analysis.
Since \texttt{npm} ecosystem policies allows unrestricted access, pre-processing was a quality filter to remove invalid packages.  

\begin{lstlisting}[language=xml,
caption={Snippet from the \textit{package.json}. Note that lines 14 and 15 correspond to the entry point file that is accessible for a client user.},
label=code:package.json]
{
"name": "user-home",
"version": "2.0.0",
"description": "Get path to user home directory",
"license": "MIT",
"repository": "sindresorhus/user-home",
"author": {
"name": "Sindre Sorhus",
"email": "sindresorhus@gmail.com",
"url": "sindresorhus.com"
},
....
},
"main": [
"index.js"
],
....
"dependencies": {
"os-homedir": "^1.0.0"
},
...
}\end{lstlisting}

\begin{figure*}
	\centering
	\subfigure[JavaScript libraries grouped by FuncCount
	]{\label{fig:fc}
		\includegraphics[width=.9\columnwidth]{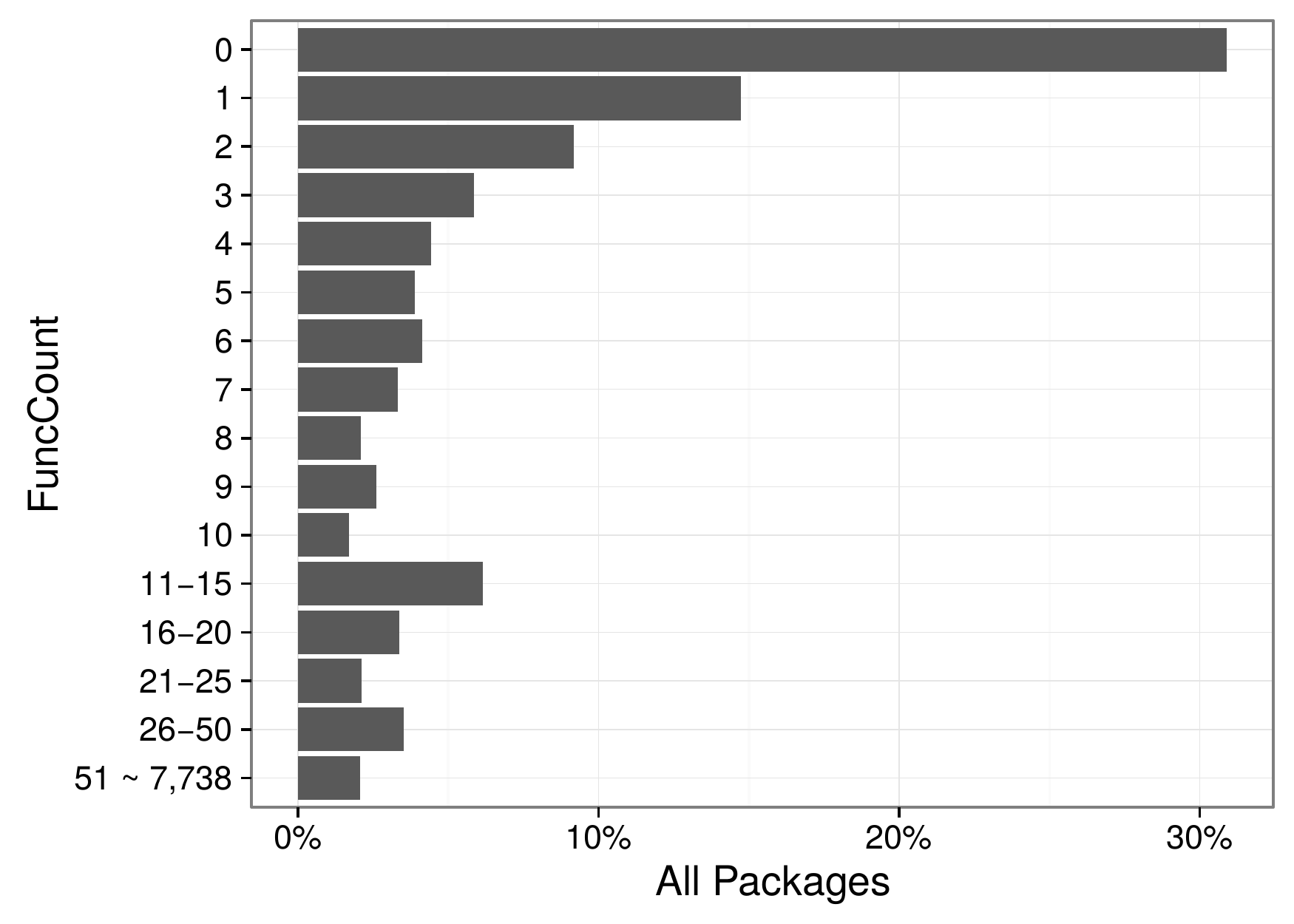}
	}
	\subfigure[Comparing Lines of Code (sLoC) between micro-packages and other packages.
	]{\label{fig:loc}
		\includegraphics[width=.9\columnwidth]{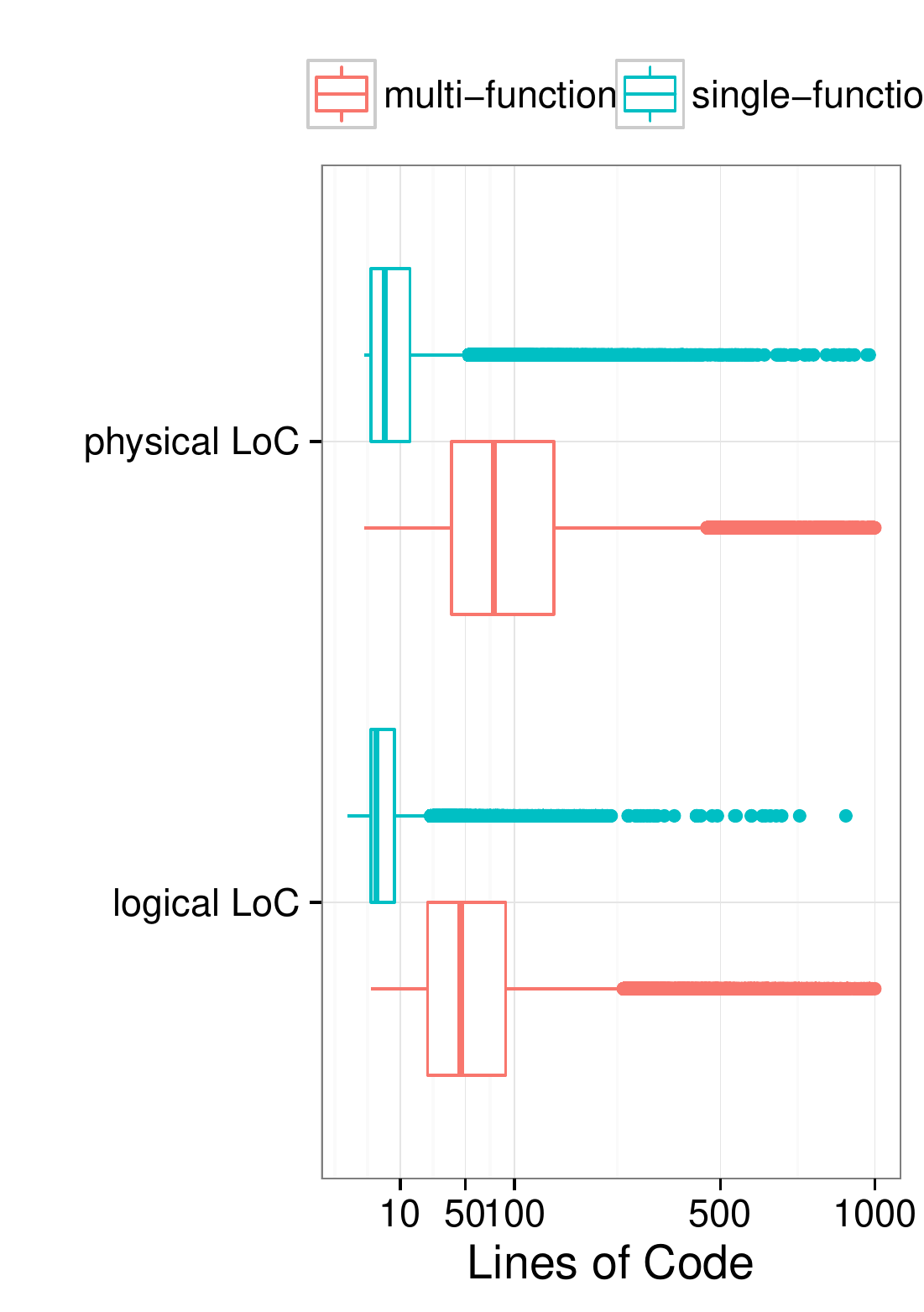}
s	}	
	\caption{Results for RQ1 package size metrics (a) FuncCount and (b) sLoC}
	\label{fig:mod}
\end{figure*}

Listing \ref{code:package.json} depicts the \texttt{package.json} metadata file used to identify and extract the \texttt{entry point file}.
This entry point will determine the available API of a package.
From this metafile, we extract the main field (Lines 14 and 15), which identifies the entry point.
Note that the \texttt{name} attribute is a unique identifier for any library. 
We then use the \texttt{esprima} npm package \cite{NPM:esprima} to extract our package size metrics.
\texttt{esprima} is a high performance  and highly popular\footnote{as of Feb 10, recorded over 17,700,000 downloads from the npm} JavaScript analysis tool that can construct a JavaScript syntax tree and loads all related class files.

\begin{table}
	\begin{center}
		\caption{Library Design Size Metric Statistics for RQ1
		}
		\label{tab:RQ1Stats}
		\begin{tabular}{lrrrrrc}
			\hline
			\multirow{1}{*}{} 
			& \multirow{1}{*}{\textbf{Min.}}
			& \multirow{1}{*}{\textbf{Median ($\bar{x})$}}
			& \multirow{1}{*}{\textbf{Mean ($\mu$)}}
			& \multirow{1}{*}{\textbf{Max.}}\\
			FuncCount  	& 0.00 &  2.00 &   8.65  & 7,738.00\\ 
			physical SLoC & 1.00  &  29.00 &    112.30 &  137,700.00   \\
			logical SLoC &0.00 &   17.00 &    63.63  & 129,100.00 \\\hline
		\end{tabular}
	\end{center}
\end{table}

\subsubsection{\underline{Findings}}
Table \ref{tab:RQ1Stats} and Figure \ref{fig:mod} report the results of RQ1. 
Table \ref{tab:RQ1Stats} shows the summary of statistics our package size metrics, while the figure is a visual distribution of the funcCount and sLoC metrics.

From a topological viewpoint, we observe from Table \ref{tab:RQ1Stats} that libraries have a median of two functions per library. 
The result shows that most \texttt{npm} packages contain a small number of API functions available from the entry point file. 
Furthermore, we observe that the median size of the source code is between 17 loc (logical) and 29 loc (physical).
Looking at the largest package, we find that the largest in terms of both API and lines of code is the  (\texttt{tesseract.js-core})\footnote{Size of the code can be summarized by the description \textit{`if you're a big fan of manual memory management and slow, blocking computation, then you're at the right place!'}  \url{https://github.com/naptha/tesseract.js-core}} package. 
This package contains 5,951 functions and 137,700 lines of code. 


A key observation from Figure \ref{fig:fc} is that micro-packages account for up to 47\% of all \texttt{npm} libraries. 
We can observe from the figure that 32\% of all packages are without function (\ie~$FuncCount=0$).
Our manual validation shows that many of these packages belong to the \texttt{facade} micro-package type.
However, some of these micro-packages are used to only local modules so are not bridging APIs to other external packages.
For example, the \texttt{12env} package contains only the single line: 

\begin{lstlisting}[language=Python,
label=code:user-home]
module.exports = require('./lib/config')();\end{lstlisting}

that is tasked to load the internal \texttt{config.js} module that performs the computation of this package.
Further analysis reveals \texttt{config.js} to have a single API function with 71 lines of code.
Although the config module is part of the package, we conclude according to our micro-package definitions it still  has an excessive interoperability dependency.  


Figure \ref{fig:loc} shows a statistical measure of the physical sizes of micro-packages compared to the rest of the npm packages. 
We observe that micro-packages are statistically smaller in size. 
We report micro-packages have a lower median  of lines of code (physical ($\bar{x}=5 LoC$) and logical($\bar{x}=3 LoC$))
This result is not surprising and matches with our intuitions of micro-packages. 
From the figure, we also observe that not all micro-packages contain small code. 
For instance, the \texttt{icao}\footnote{is a library that looks up hard-coded International Civil Aviation Organization airport codes. The website is at \url{https://github.com/KenanY/icao/blob/master/index.js}} package has 8,255 lines of code.
Further analysis concludes that most of the code include hard-coded variable constants.

\begin{hassanbox}
In summary for RQ1, we propose our package size metrics. To answer RQ1: \textit{ our findings show that micro-packages account for a significant 47\% of all packages, with 32\% acting as a facade to create excessive dependencies within the ecosystem. Interestingly, most \texttt{npm} packages are designed to be smaller in size, with up to 2 functions and 29 lines of code per package. 	 
}
\end{hassanbox}

\subsection{\textbf{RQ2: \RqTwo}}
\subsubsection{\underline{Motivation}}

Our motivation for RQ2 is to understand the extent of how micro-packages form dependency chains across the ecosystem. 
In this research question, we calculate the dependency chains of \texttt{npm} packages to survey whether or not micro-packages result in an increase of dependencies.
Therefore, we investigate whether or not there is a difference in dependency complexity for micro-packages.
RQ2 includes testing the hypothesis that \textit{micro-packages include lesser dependency complexity than other packages in the npm ecosystem}.

%

\begin{figure}
	\centering
	\includegraphics[width=0.5\textwidth]{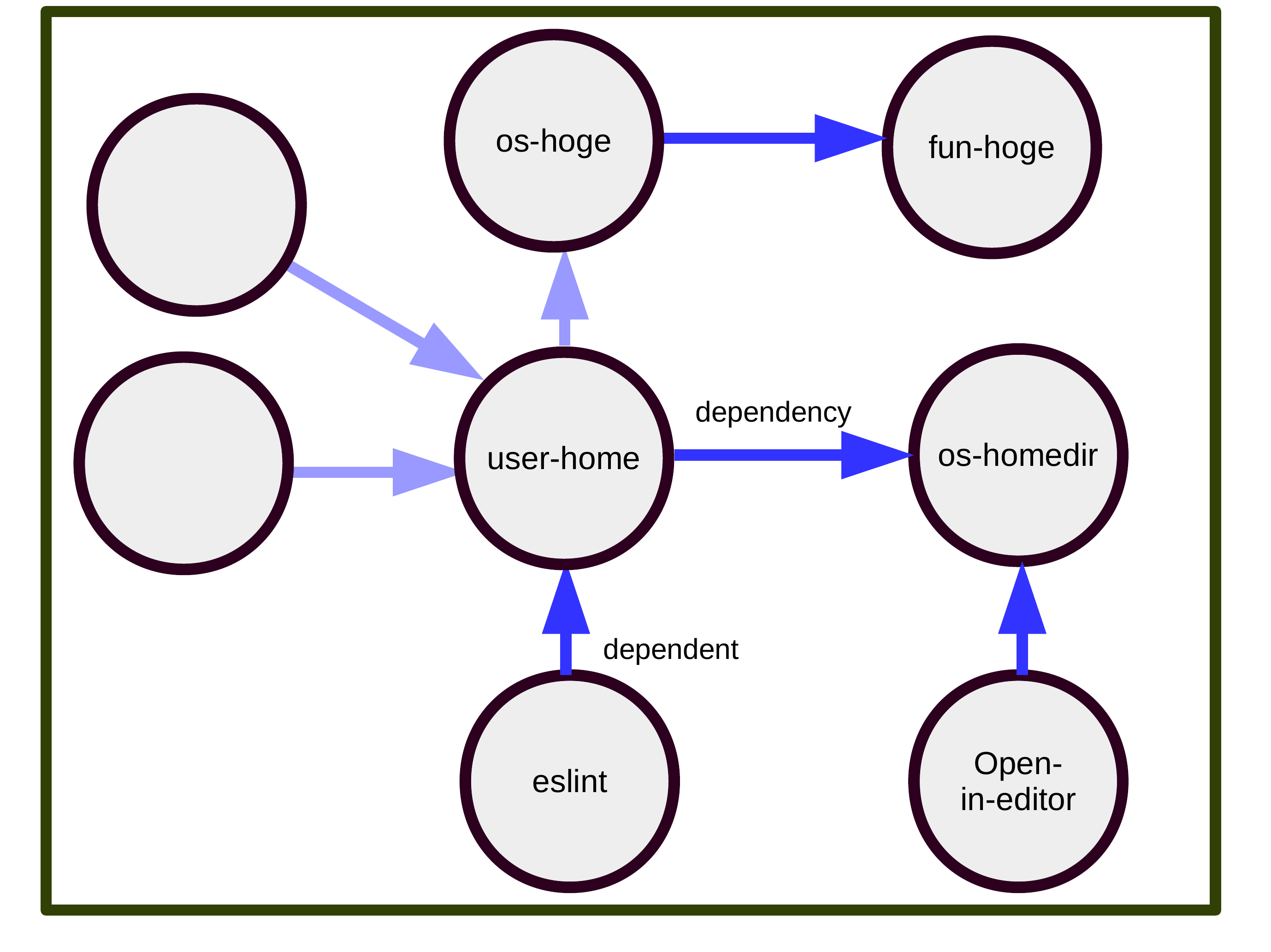}
	\caption{Modeling a chain of dependencies for the \texttt{eslint} package as a directed acyclic graph network. For better examples, we added \texttt{os-hoge} and \texttt{fun-hoge} libraries}
	\label{fig:SUG}
\end{figure}

\subsubsection{\underline{Research Method}}
Our research method to answer RQ2 is by using network analysis\cite{wasserman1994social} with statistical analysis of the network generated metrics and then manual validation to understand our results.
It is performed in two steps.
The first step involves the construction of \texttt{npm} network that models and represents dependency chains in the \texttt{npm} ecosystem.

The second step involves analysis of the generated \texttt{npm} graph network.
To do this, we propose a set of dependency metrics derived from the network analysis to describe a npm package dependency complexity.
We then utilize box-plots and violin plots with each dependency metric to depict the differences between micro-packages to the rest of the packages.
Finally, in our analysis we test the hypothesis whether or not micro-package has greater or lesser dependency complexity that other packages. 
To assess the significance of dependency metric differences, we use the Wilcoxon–Mann–Whitney and Cohen's \textit{d} test  \cite{Romano:2006} as they do not require the assumption of normal distribution. 
The null hypothesis ($H_0$) states that \textit{either micro-packages include greater or lesser dependency complexity}\footnote{We set the confidence limit $\alpha$ as 0.01} 
If the null hypothesis is accepted, we then determine which population has the higher means to state our hypothesis. The alternate hypothesis  ($H_1$) is that \textit{micro-packages include identical dependency complexities to other packages.} 
Furthermore, to assess the difference magnitude, we studied the effect size based on Cohen's \textit{d}. The effect size is considered: (1) small if 0.2 $\leqslant$ \textit{d} $<$ 0.5, (2) medium if 0.5 $\leqslant$ \textit{d} $<$ 0.8, or (3) large if \textit{d} $\geqslant$ 0.8.

\subsubsection{Modeling the \texttt{npm} ecosystem network}
Figure \ref{fig:SUG} serves as our graph-based model of dependencies in the \texttt{npm} ecosystem. 
In this model, we depict graph nodes as each a unique package and edges as directed dependencies. 
Formally, we define our \texttt{npm} ecosystem network as a directed graph $G = (V, E)$ where $V$ is a set of nodes (=packages) and E $\subseteq$ $V\times V$ is a set of edges (=dependencies). 
Since the graph is directed, $E$ has an ordering: $\{u,v\}\neq\{v,u\}$. 
For instance, we represent the \texttt{eslint} package dependency on \texttt{user-home} package. 
We observe that the \texttt{eslint} package is transitively dependent on \texttt{os-homedir} through its dependence on the \texttt{user-home} package. 

Furthermore, we define dependency metrics to describe the direct and transitive dependencies.
We first introduce the following direct dependency metrics:
\begin{itemize}
	\item \textit{Dependents (Incoming Degree)} - is the number of packages that directly depend on a package (incoming dependencies). 
	In figure \ref{fig:SUG} \texttt{user-home} has three dependent (\ie~includes the \texttt{eslint} package).
	Dependents is a useful measure of package reuse within the ecosystem.
	
	\item \textit{Dependencies (Outgoing Degree)} - is the number of packages on which a package directly depends on (outgoing dependencies). 
	In figure \ref{fig:SUG} \texttt{user-home} has two dependencies (\ie~\texttt{os-homedir} and \texttt{os-hoge}).
	
	
\end{itemize}

We then introduce our transitive dependency metrics:


\begin{itemize}	
	\item \textit{Chain Length (Eccentricity score)} - is a normalized measure to evaluate the longest chain of dependencies for a library dependency. 
	This metric counts the longest path for a given graph node to any of its reachable nodes. 
	As shown in figure \ref{fig:SUG}, \texttt{eslint} has a chain length of three (\ie~$\textit{eslint} \rightarrow \textit{user-home} \rightarrow \textit{os-hoge} \rightarrow \textit{fun-hoge}$).
	
	
	\item \textit{Reach Dependencies ($\#$ of reachable nodes)} - is the number of libraries that are within its reachable path. 
	For library \texttt{eslint} has four reachable nodes (\ie~$\{$\texttt{user-home}, \texttt{os-homedir}, \texttt{os-hoge} and \texttt{fun-hoge}$\}$).	
\end{itemize}
\label{sec:motive2}

\subsubsection{\underline{Dataset}}
	\begin{lstlisting}[language=xml,
	caption={Snippet from the \textit{package.json} of  the \texttt{user-home} in RQ1 (See Section \ref{sec:rm1}) that shows the header and dependency relation to \texttt{os-homedir} library.},
	label=code:snippackagen]
	{
	"name": "user-home",
	"version": "2.0.0",
	....
	"dependencies": {
	"os-homedir": "^1.0.0"
	},
	...
	}\end{lstlisting}

As shown in Listing \ref{code:snippackagen}, we extract each dependency (\ie~Lines 5 and 6.) from the \texttt{package.json} metadata file to generate our npm graph network.  
For each package, we collect all runtime dependencies, development and optional dependencies. 

\begin{algorithm}
	\small
	\caption{\texttt{npm} Ecosystem Network Algorithm}\label{alg:genGraph}
	\KwData{$V$ is a package graph node
		$E$ is a dependency edge,\\
		$packageList$ lists all collected packages,\\
		$getDeps(package)$ list all dependencies for a package lib\\
		... \\      
	}
	
	Initialize $packageList$ with all packages\;
	Initialize npmGraph $G$\;
	
	\For{ \textbf{each} package \textbf{in} packageList}{
		\eIf{pacakge lib \textbf{in} $G$}{
			node $V_{lib} \gets getNpmGraphNode(G,lib)$\;
		}{
		create node $V_{lib}\gets createNode(lib)$\;
		add node $V_{lib}$ to $G$\;	
	}
	dependencyList $\gets$ getDeps(lib)\;
	\For{ \textbf{each} library dep \textbf{in} dependencyList}{
		create node $V_{dep} \gets createNode(dep)$\;
		add node $V_{dep}$ to $G$\;
		create edge $E$ $\gets$ $createEdge(V_{dep},V_{lib})$\;
		add edge $E$ to $G$\;
	}
}
\end{algorithm}

Algorithm \ref{alg:genGraph} details the algorithm we use to construct and generate the \texttt{npm} graph network. 
The key idea in building the \textit{npm} network is to append each package as a new graph node to existing packages in the network.
In this sense, a node is a package with that is either dependent or depended upon by other package nodes in the network.
So for each new package, we first determine if it exists in the graph (Step 4) as a depended node from another package. 
If its does not exist (Steps 7-8), we then proceed to create a new node for this package and then check its dependent whether they also exist on the graph (Steps 11-15).

\begin{figure*}
	\centering
	\subfigure[Direct Dependents
	]{\label{fig:DepIn}
		\includegraphics[width=0.45\columnwidth]{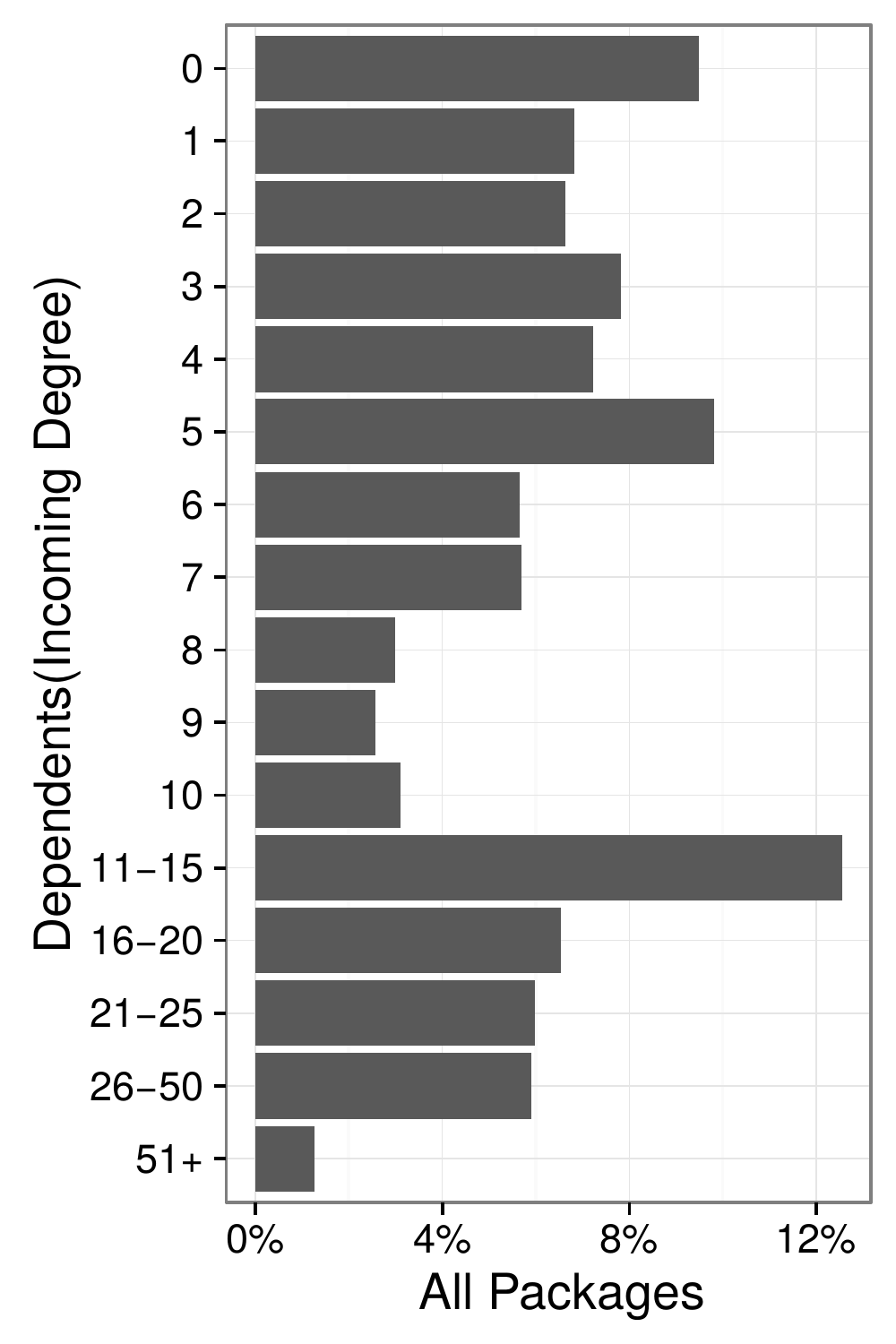}
	}
	\subfigure[Direct Dependencies
	]{\label{fig:DepOut}
		\includegraphics[width=0.45\columnwidth]{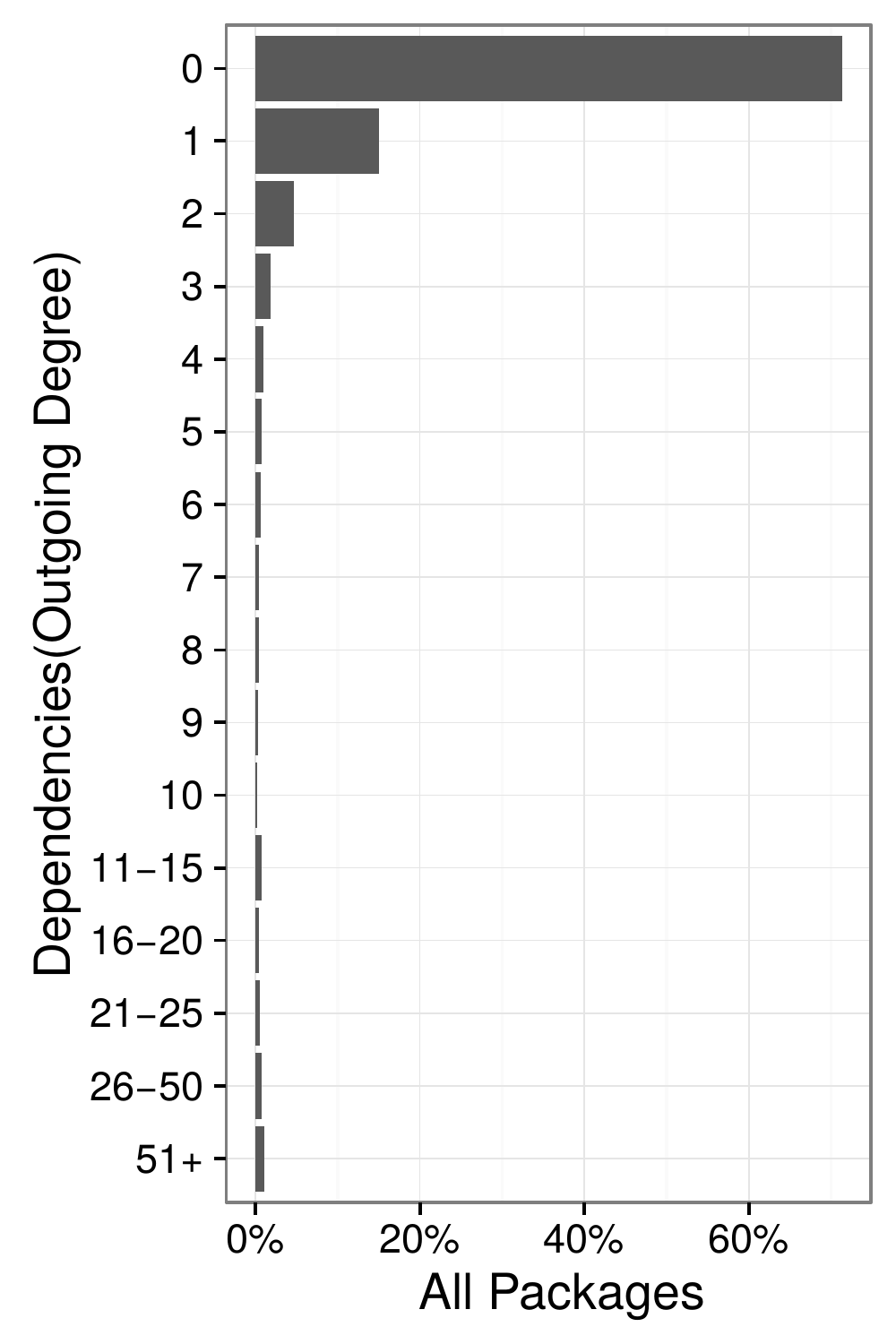}
	}
	\subfigure[Transitive Chain Length
	]{\label{fig:ecc}
		\includegraphics[width=0.45\columnwidth]{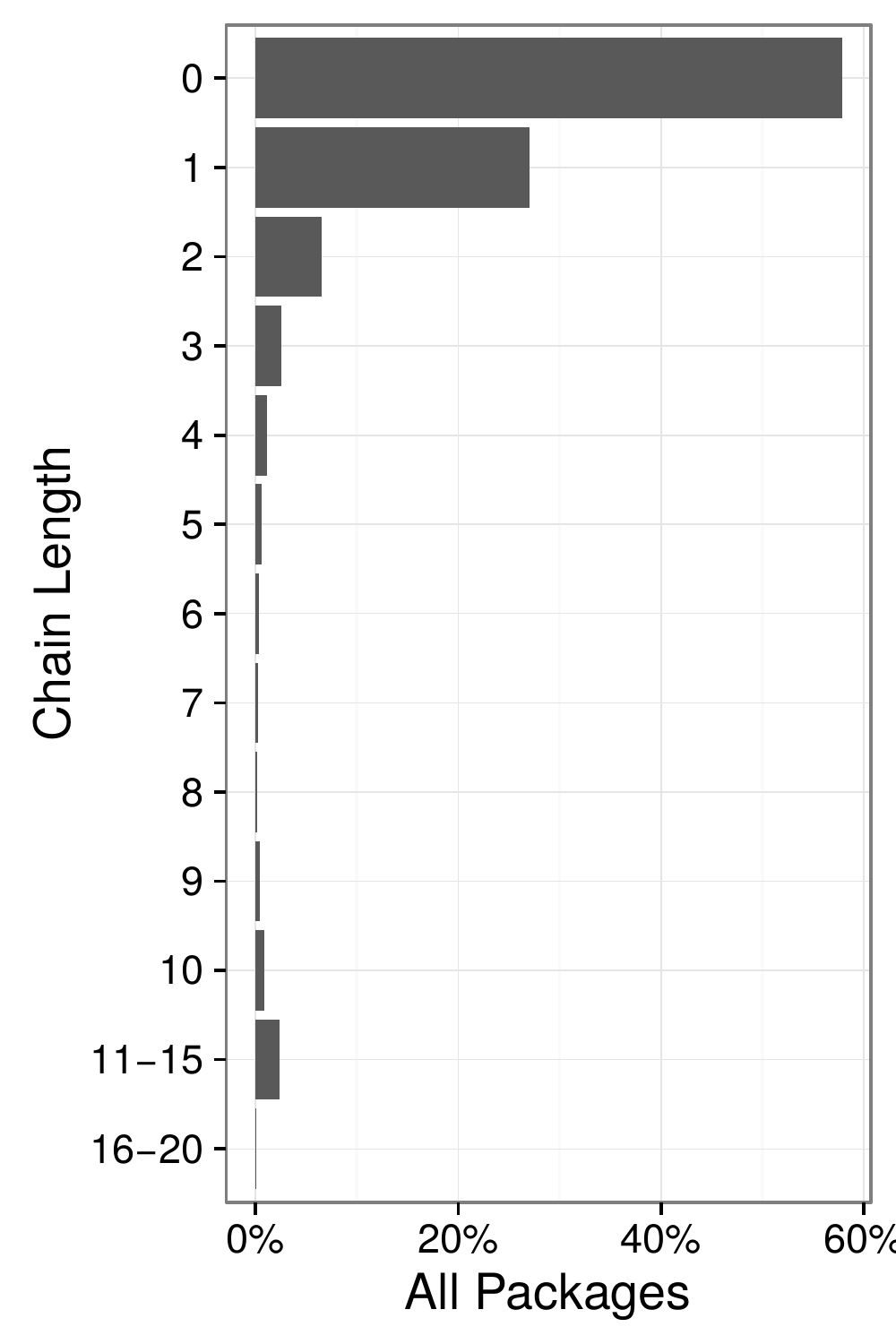}
	}
	\subfigure[Transitive Reach Dependencies
	]{\label{fig:rd}
		\includegraphics[width=0.45\columnwidth]{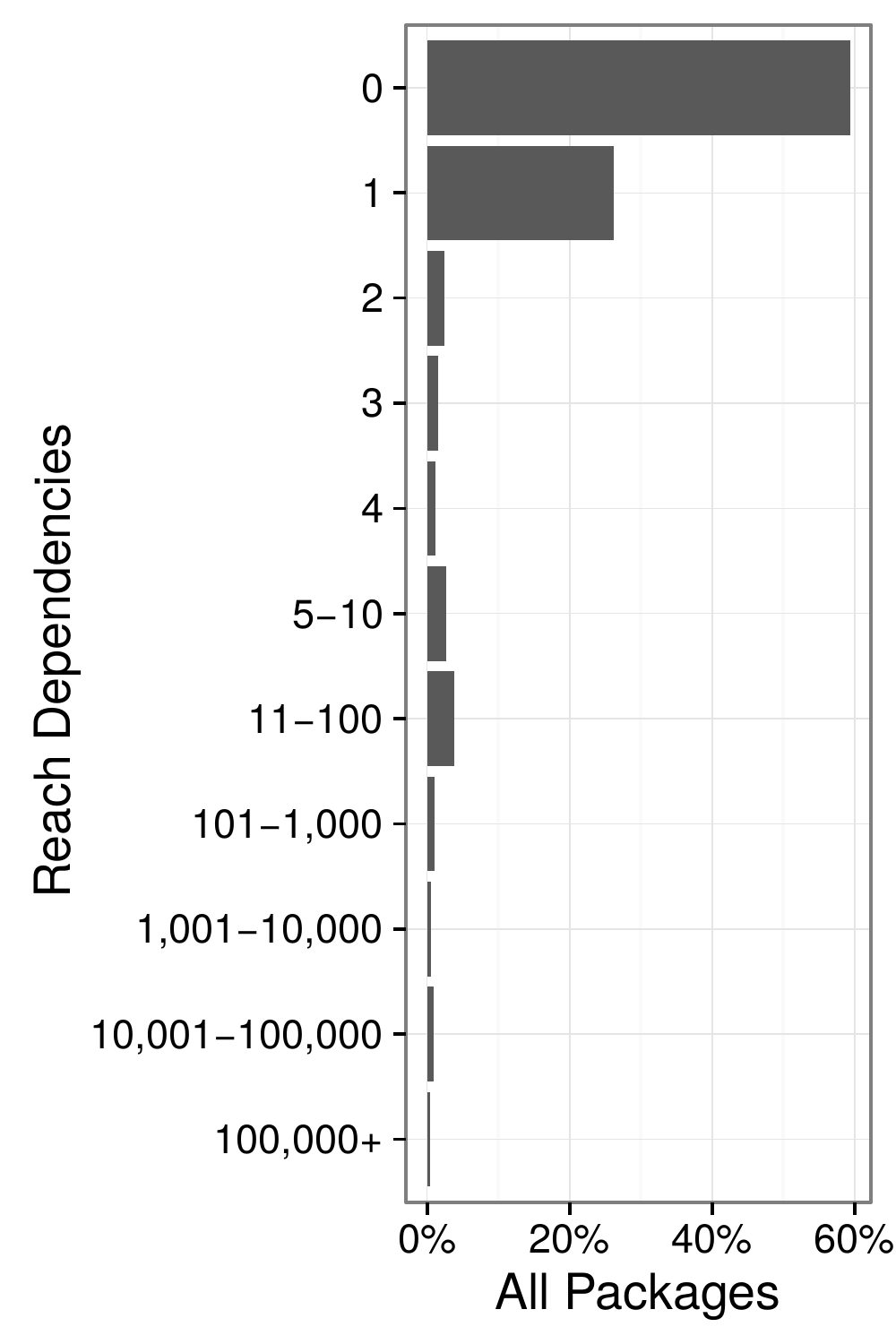}
	}	
	\caption{Results for RQ2 direct and transitive dependency metrics for the \texttt{npm} ecosystem network}
	\label{fig:allDep}
\end{figure*}

\begin{table}
	\begin{center}
		\caption{Summary Statistics of the \texttt{npm} ecosystem network for RQ2
		}
		\label{tab:datasetRQ2}
		\begin{tabular}{lcc}
			\hline
			\multirow{1}{*}{} 
			& \multirow{1}{*}{\textbf{\texttt{npm} ecosystem network}}
			\\
			\# nodes&169,964&\\ 
			\# edges&986,075& \\ 
			Average Chain Length & 5.89& \\ \hline
		\end{tabular}
	\end{center}
\end{table}

Table \ref{tab:datasetRQ2} shows that our \texttt{npm} graph network comprises of 169,964 graph nodes and 986,075 graph edges with an average chain length of 5.89. 
We use python scripts to implement the algorithm and the Neo4j graph database to store the npm network (\ie~\textit{py2neo} \cite{neo4j}, the \textit{neo4j}  \cite{neo4j}.
The \textit{gephi} tool \cite{gephi} was used to generate the dependents, dependencies and chain length dependency metrics.
Furthermore, we used a Depth First Search (DFS) query in Neo4j to generate the Reach Dependencies metric. 
For a package node $v$ with a package name $v.name$, we use the Neo4j cypher query\footnote{For realistic computation-time costs, we set a threshold of our reachable node to four chains of dependencies (x=5). As the average chain length is 5.89 it should have minimal effect on the result (See Table \ref{tab:datasetRQ2})}: 
	\texttt{Match v$\{$name:v.name$\}$ -[*..x]->(w) where NOT v-w RETURN v.name, count(DISTINCT(v))}
to find all packages that are within its dependency reach in the ecosystem.

\begin{table}
	\begin{center}
		\caption{Dependency metric statistics of the collected \texttt{npm} for RQ2
		}
		\label{tab:RQ2Ego}
		\begin{tabular}{lrrrrrc}
			\hline
			\multirow{1}{*}{} 
			& \multirow{1}{*}{\textbf{Min.}}
			& \multirow{1}{*}{\textbf{Median ($\bar{x})$}}
			& \multirow{1}{*}{\textbf{Mean ($\mu$)}}
			& \multirow{1}{*}{\textbf{Max.}}\\
			Dependents 	&  0.00 &     3.00  &   5.81 & 22,750.00 \\ 
			Dependencies &  0.00 &     0.00   &  5.80 & 41,550.00   \\
			Chain Length&  0.00 &  0.00 &  1.03 & 20.00 \\ 
			Reach Dependents & 0.00  &  0.00 &    563.90 &  121,600.0    \\\hline
		\end{tabular}
	\end{center}
\end{table}

\begin{figure*}
	\centering
	\subfigure[Direct Dependents
	]{\label{fig:DeppComp}
		\includegraphics[width=0.4\columnwidth]{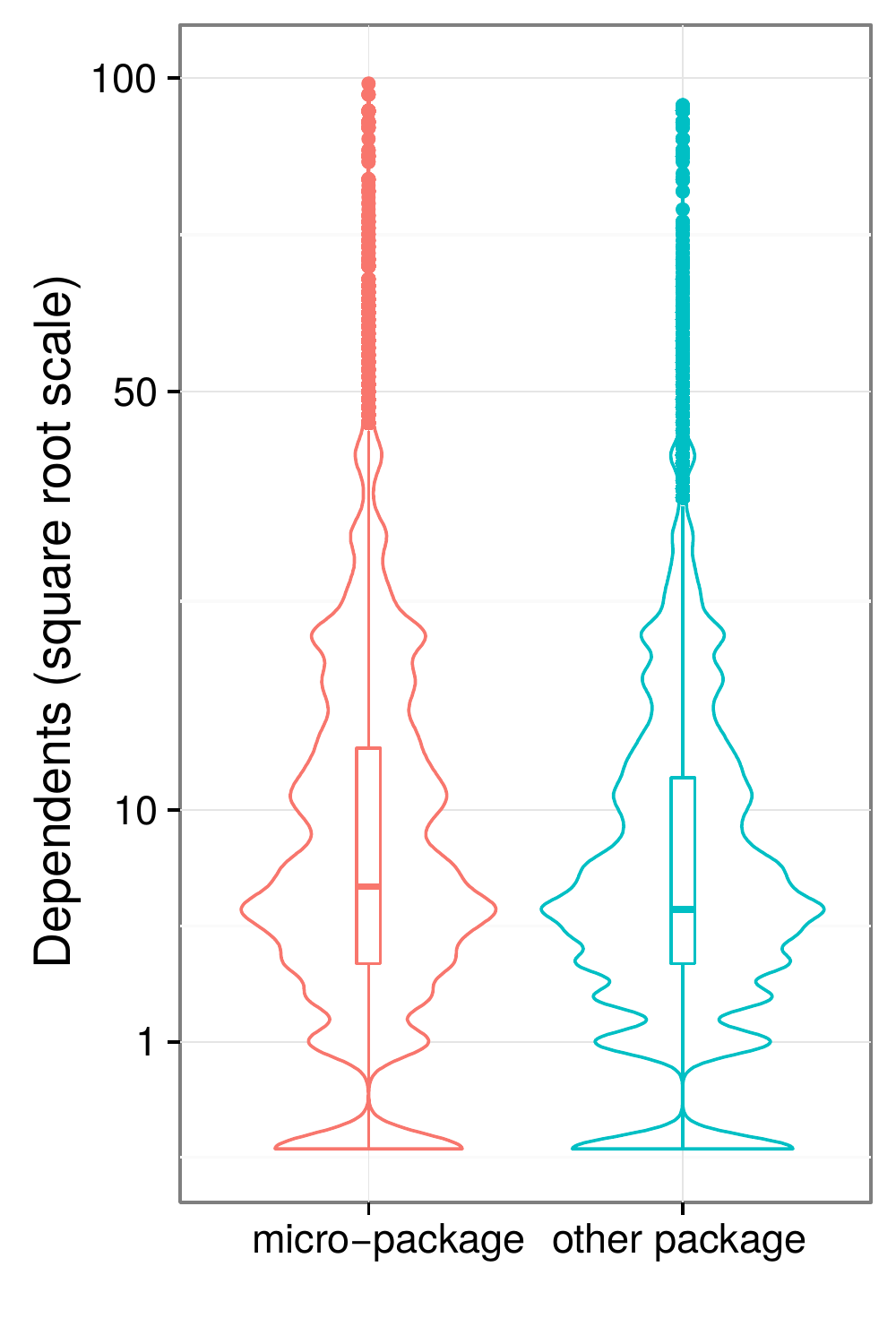}
	}
	\subfigure[Direct Dependencies
	]{\label{fig:DepComp}
		\includegraphics[width=0.4\columnwidth]{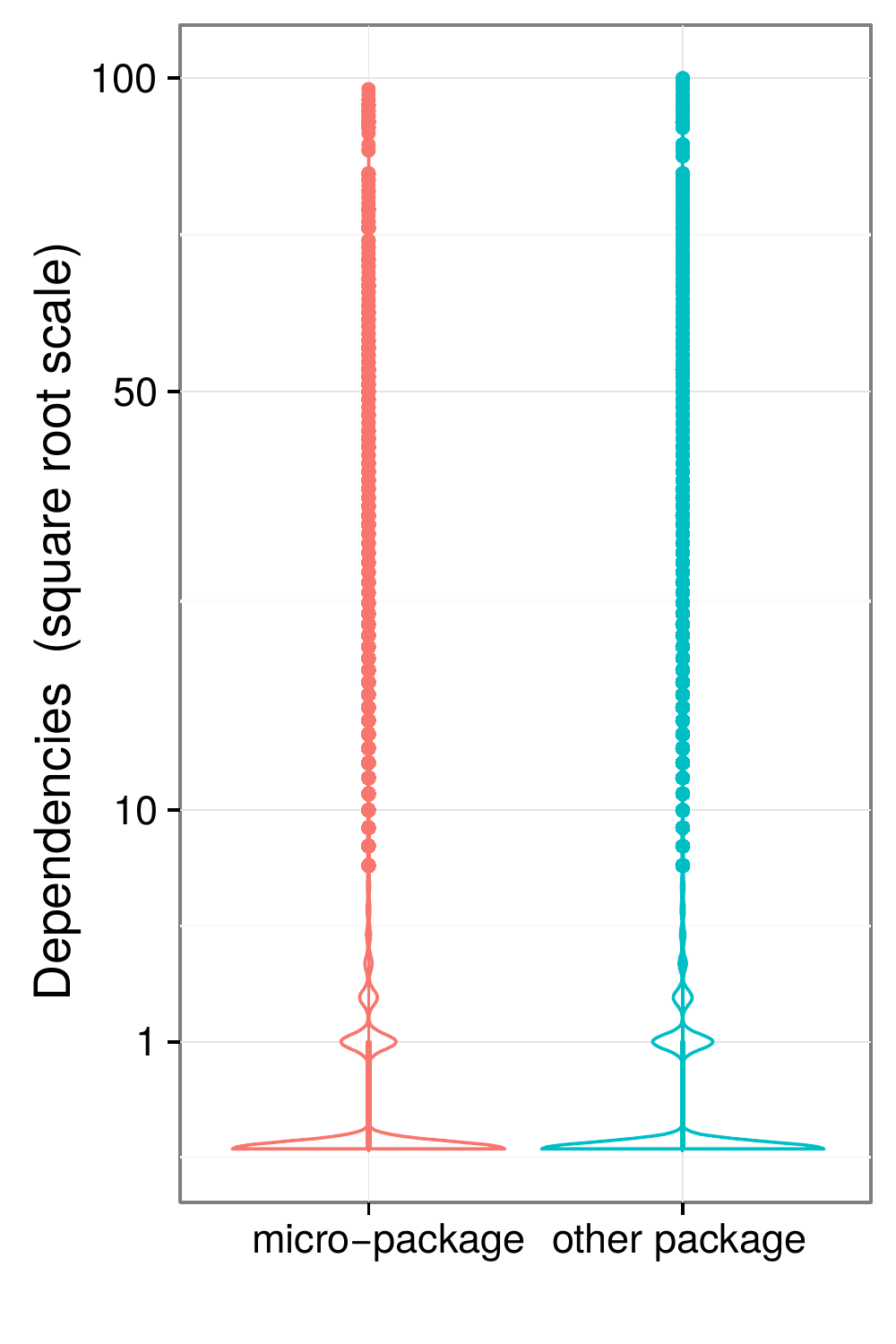}
	}
	\subfigure[Transitive Chain Length
	]{\label{fig:EccComp}
		\includegraphics[width=0.4\columnwidth]{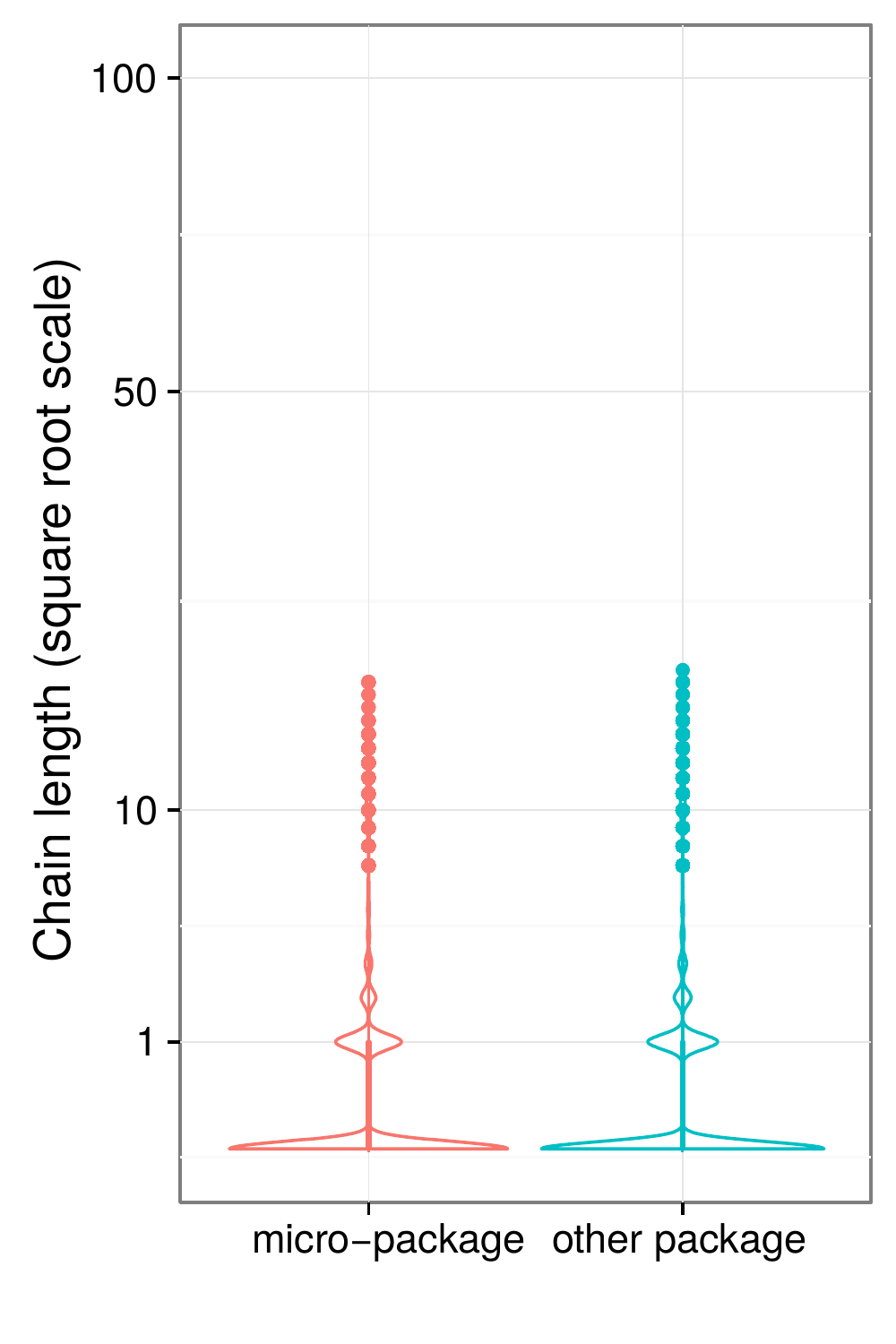}
	}
	\subfigure[Transitive Reach Dependencies
	]{\label{fig:RDComp}
		\includegraphics[width=0.4\columnwidth]{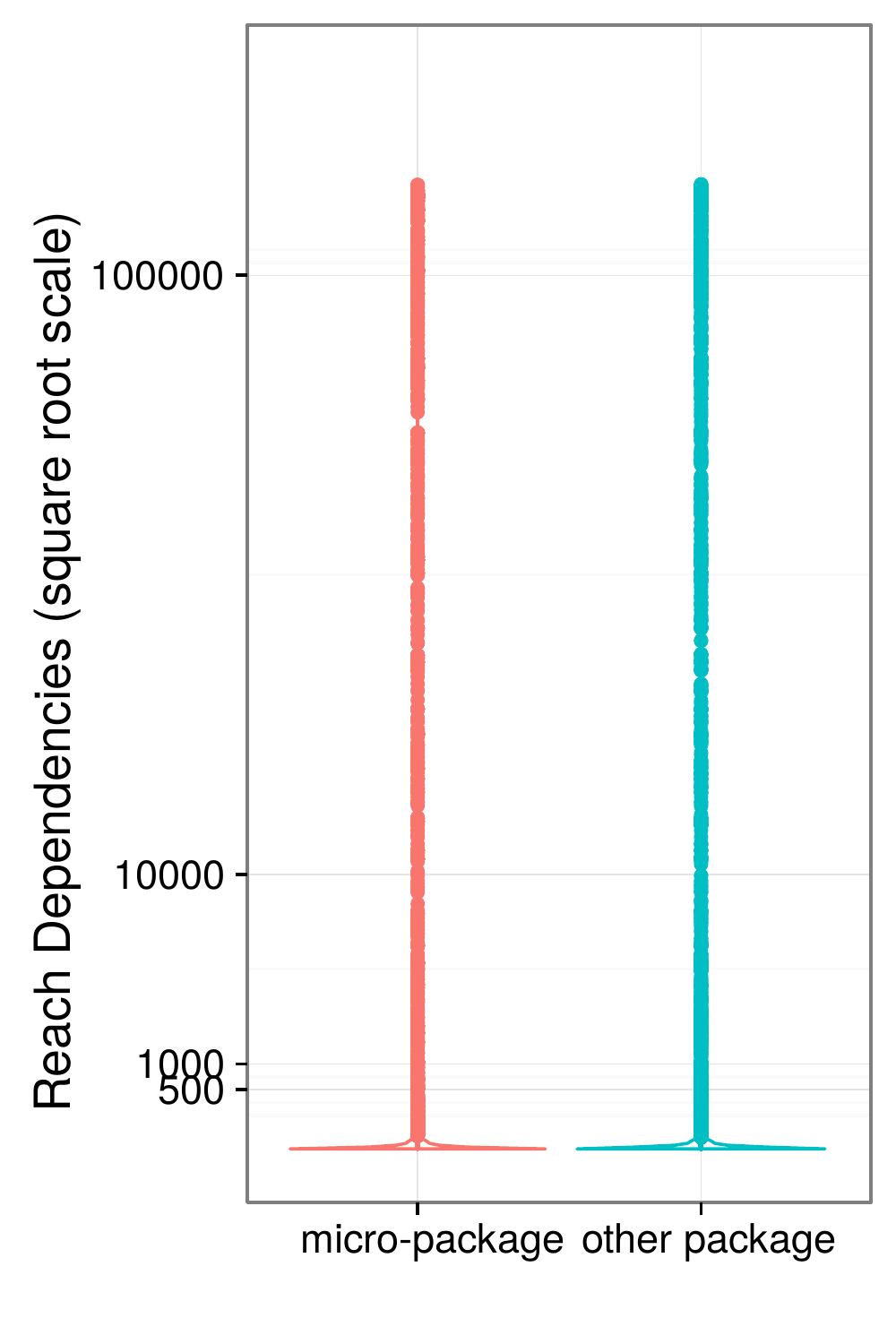}
	}
	\caption{Micro-package dependency metric comparisons against the other packages in the \texttt{npm} ecosystem network.}
	\label{fig:RQ2Comp}
\end{figure*}

\subsubsection{\underline{Findings}}
Table \ref{tab:RQ2Ego} shows the summary statistics for the dependency metrics generated for the \texttt{npm} network.
From a topological viewpoint, we observe from that most packages had a median of three dependents per with no dependencies. 
With a manual validation, we find that the \texttt{everything} package has the most dependents in the network.
Interestingly, this package is labeled by many JS developers as a \textit{`hoarder'}\footnote{blog discussing this at \url{https://github.com/jfhbrook/hoarders/issues/2}} library.
With over 22,750 other incoming dependency relations from other libraries, this package offer no useful function but to access every other package on the network. 
Conversely, we observe the very useful \texttt{mocha}\footnote{website at \url{https://github.com/mochajs/mocha}} package as the most depended upon package (\ie~Dependencies = 41,570). This testing framework validates the notion that \texttt{npm} library developers indeed run and perform test on their npm packages.

Figure \ref{fig:allDep} shows the statistical distribution for all direct dependency metrics across the \texttt{npm} ecosystem. 
The key observation from Figure \ref{fig:DepIn} is that only 9\% of all packages have no dependents  (\ie Dependents=0), inversely indicating that almost 91\% of libraries are being dependent upon by another package in the ecosystem.
Findings also provide evidence that more than 12\% of the libraries have about 11 to 15 dependents. 
Figure \ref{fig:DepOut} confirms the results in Table \ref{tab:RQ2Ego}, that clearly showing that 65\% of packages in the npm ecosystem show no dependencies to other packages. 

Figure \ref{fig:ecc} and Figure \ref{fig:rd} we show through our transitive dependency metrics that around 60\% of packages are not transitively reachable from another package.
With a manual validation we find two packages that are critical to the \texttt{npm} ecosystem.
The network analysis shows that \texttt{fs-walk}\footnote{website at \url{https://github.com/confcompass/fs-walk}} having the longest transitive dependency chain length (\ie~Chain length = 20). 
This useful library is used for synchronous and asynchronous recursive directory browsing. 
Its long dependency chain provides evident that it is a critical package within the \texttt{npm} ecosystem as it is part of multiple dependency chains. 
We also report the very popular and useful AST-based pattern checker tool \texttt{eslint}\footnote{\url{https://www.npmjs.com/package/eslint}} to be the most reachable by any other package on the \texttt{npm} ecosystem.  

Testing our hypothesis that \textit{either micro-pacakage include greater or lesser dependency complexity} shows that  dependency complexities of a micro-package is not so trivial, with some micro-packages have just as long chain lengths and reach dependencies as the rest of the packages. 
Although visually in Figure \ref{fig:RQ2Comp} there seems no transitive dependency differences between micro-packages and the rest of packages, results of the Wilcoxon–Mann–Whitney tests show otherwise.
Table \ref{tab:RQ2TComp}  show that a micro-package are more likely to be reused, indicated by statistically more dependents than the rest of the packages (\ie~Dependent accepted H$_0$ for micro-package).  
Furthermore, the results show micro-packages are more likely to reach other packages (\ie~Reach Dependencies accepted H$_0$ for micro-packages) in the \texttt{npm} ecosystem. 


\begin{table}
	\begin{center}
		\caption{Statistical Test results of Dependency Metrics for micro-packages against the rest of the packages. The ($>\mu$) column shows the library type with the greater means between the populations.  
		}
		\label{tab:RQ2TComp}
		\begin{tabular}{llccrrc}
			\hline
			& \multirow{1}{*}{\textbf{$H_0$}}
			& \multirow{1}{*}{\textbf{$>\mu$}}
			& \multirow{1}{*}{\textbf{p-value}}
			& \multirow{1}{*}{\textbf{\textit{d}}}\\
			Dependents 	&  \colorbox{green!25} {accept}& micro-package & $<0.01$  & 0.09 (S)  \\ 
			Dependencies & \colorbox{red!25} {reject} & identical&$<0.13$ & 0.13 (S) \\
			Chain Length& \colorbox{green!25} {accept} & other packages &$<0.001$  & 0.01 (S)    \\
			Reach Dep.& \colorbox{green!25}{accept} & micro-package&$<0.002$ &0.01 (S) \\\hline
		\end{tabular}
	\end{center}
\end{table}

\begin{hassanbox}
In summary for RQ2, we propose and model a \textit{npm} dependency network for the \texttt{npm} ecosystem.
To answer RQ2: \textit{ our findings depict micro-packages dependencies as not so trivial, with some micro-packages having just as long dependency chain lengths and reachability to other packages in the \texttt{npm} ecosystem.} 
\end{hassanbox}

\subsection{\textbf{RQ3: \RqThree }}

\subsubsection{\underline{Motivation}}
\label{sec:RQ3m}
Our motivation for RQ3 is to understand the whether micro-packages provide any usage benefits in terms of server-side NodeJs performance for developer. 
In a 2015 developer blog\footnote{The blog discusses how the \texttt{require()} function impacts testing time due to its slow loading into the environment: at \url{https://kev.inburke.com/kevin/node-require-is-dog-slow/}}, web developers express how the loading times affects the application performance when using JavaScript based modules on the NodeJs platform environment.  
According to the NodeJs documentation\footnote{\url{https://nodejs.org/api/modules.html\#loading_from_node_modules_Folders}} NodeJs uses a complex searching algorithm to locate the requested source code each time the \texttt{require()} function is invoked.
Therefore, we investigate whether there is a difference in the package loading times for micro-packages compared to the rest of the packages.
RQ3 includes testing the hypothesis that \textit{micro-packages incur developer usage costs of a micro library are no less than the rest of packages in the \texttt{npm} ecosystem}.

\subsubsection{\underline{Research Method}}
Our research method to answer RQ3 is by using controlled experiment simulations with statistical analysis and manual validation to understand our results.
It is performed in two steps.
The first step comprises of candidates \texttt{npm} packages eligible for the experiment.
In our preliminary experiments, encountered  packages that some malicious packages caused disruptions to our experiment environment, particularly related to malicious libraries such as rimrafall\footnote{details found at: \url{https://github.com/joaojeronimo/rimrafall}}. 
We found other documented vulnerable threats to our experiments \footnote{\url{http://blog.npmjs.org/post/141702881055/package-install-scripts-vulnerability} and a list at \url{https://snyk.io/vuln?type=npm}}.
To mitigate this, two authors manually checked the package name for keywords that raise suspicions to dubious packages.
After 8 hours of scanning the package lists, we were able to manually validate 20,000 packages based on sufficient popular downloads usage and a recognizable package name. 

For the analysis step, we describe the developer usage cost distribution using statistical metrics (\ie~Min., Mean ($\mu$), Median ($\bar{x}$), Max.). 
We then manually investigate and pull out examples of the results to qualitatively explain the reasoning behind our conclusions.
The second step involves analysis of the experiment results to test the hypothesis that incurred usage costs of a micro-package are no less than the rest of packages in the \texttt{npm} ecosystem. 
To assess the significance of dependency metric differences, we use the Wilcoxon–Mann–Whitney and Cohen's \textit{d} test  \cite{Romano:2006} as they do not require the assumption of normal distribution. 
The null hypothesis ($H_0$) states \textit{either micro-packages incur either greater or lesser developer usage costs}\footnote{We set the confidence limit $\alpha$ as 0.01}. 
The alternate hypothesis  ($H_1$) is that\textit{ both micro-packages incur identical developer usage costs compared to the rest of the packages.}
If the null hypothesis is rejected, we then conclude that micro-packages share identical usage costs the rest of the packages.
Furthermore, to assess the difference magnitude, we studied the effect size based on Cohen's \textit{d}. The effect size is considered: (1) small if 0.2 $\leqslant$ \textit{d} $<$ 0.5, (2) medium if 0.5 $\leqslant$ \textit{d} $<$ 0.8, or (3) large if \textit{d} $\geqslant$ 0.8.


%
\label{sec:data3}

\begin{algorithm}
	\small
	\caption{Developer Usage Scenario for RQ3}\label{alg:RQ3scen}
	\KwData{$packageList$ lists all collected packages,\\
	}
	
	\For{ \textbf{each} package \textbf{in} packageList}{
	clean NodeJs env.$\gets$ clean lib folder and clear cache\;
		
	Fetch Package $\gets$ url from npm repository\;
	Install Package $\gets$ into the local NodeJs\;
	Load Package $\gets$ use require () to load into local NodeJs\;

}
\end{algorithm}

\begin{table}
	\begin{center}
		\caption{Controlled Experiment Specifications for RQ3
		}
		\label{tab:dataExpRQ3}
		\begin{tabular}{lrr}
			\hline
			\multirow{1}{*}{} 
			& \multirow{1}{*}{\textbf{Dataset statistics}}
			\\
			Experiment Period&Sept-9th-2016 $\sim$ Sept-12th-2016&\\
			Download Speed &151.96 Mbps \\
			Client Machine &Intel(R)Xeon(R), 8-core@3.60GHz\\
			 &32 GB RAM\\
			Operating System& Ubuntu ver. 14. 04\\ 
			server-side JS runtime env.& \texttt{NodeJs} ver. 4.2.6\\\hline
			Collected Packages& 20,000 \texttt{npm} packages\\
			Executed Packages& 14,126 \texttt{npm} packages\\\hline
		\end{tabular}
	\end{center}
\end{table}

\begin{figure*}
	\centering
	\subfigure[Fetch Time (in milliseconds)
	]{\label{fig:ftime}
		\includegraphics[width=0.5\columnwidth]{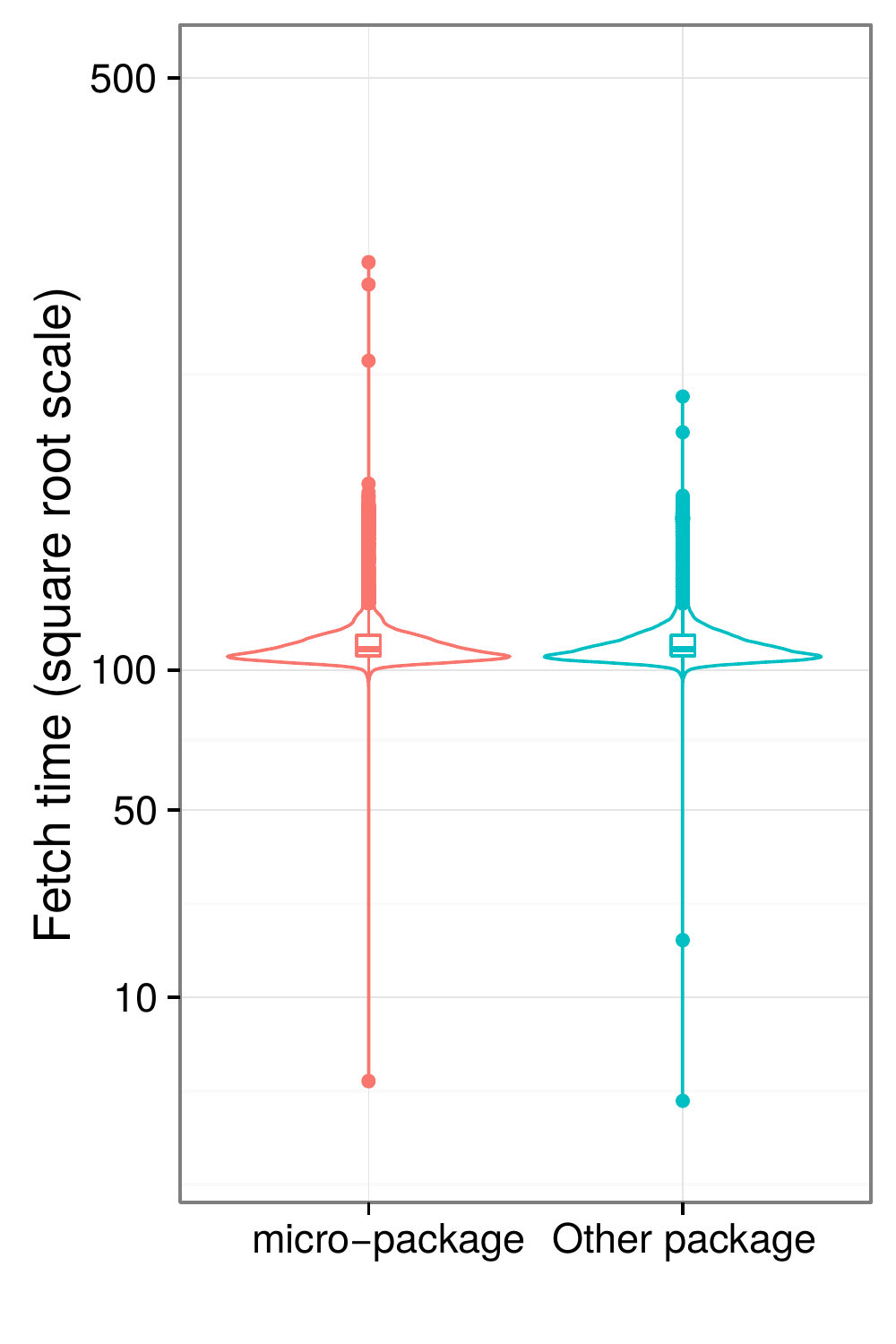}
	}	
	\subfigure[Install Time (in milliseconds)
	]{\label{fig:itime}
		\includegraphics[width=0.5\columnwidth]{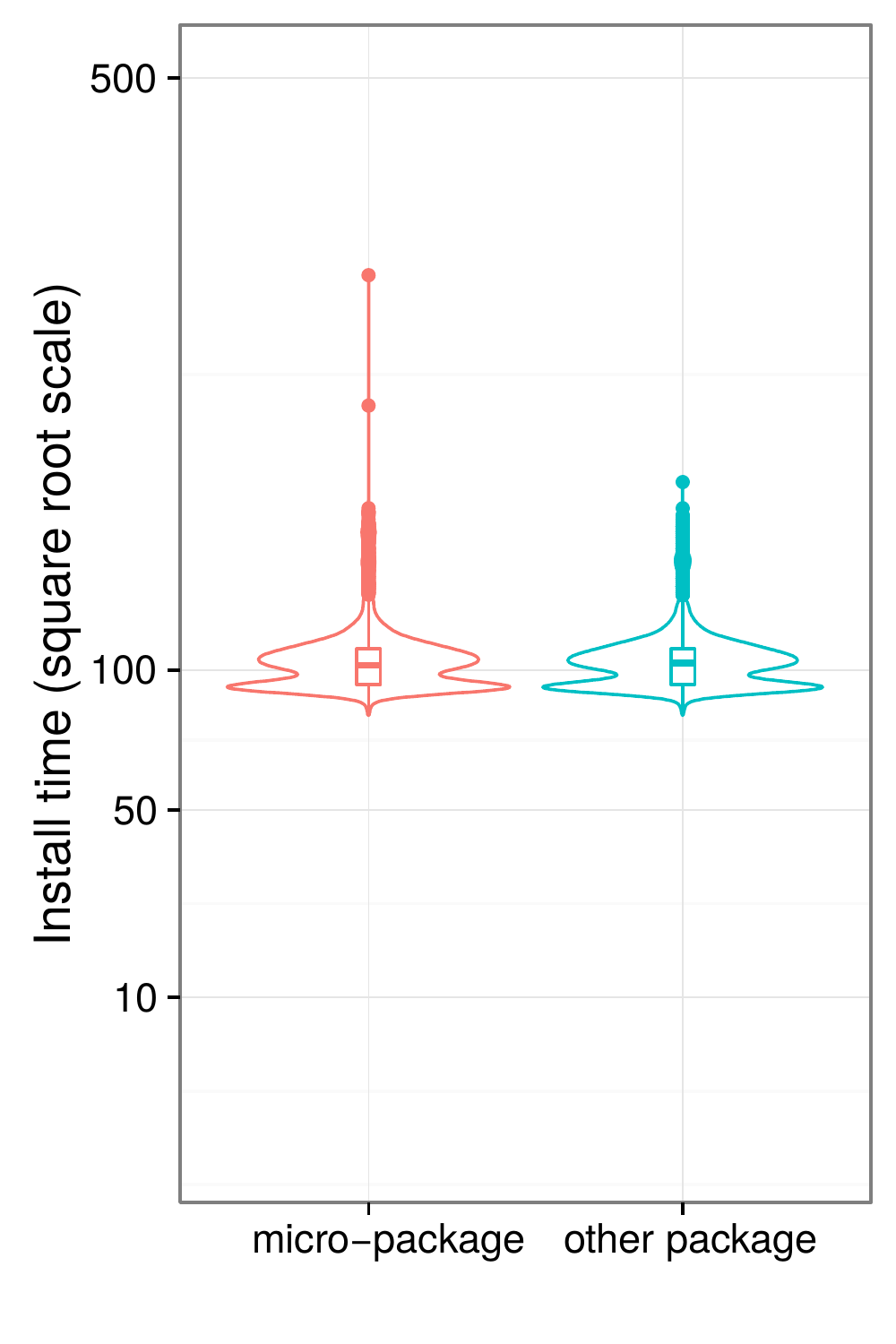}
	}
	\subfigure[Load Time (in milliseconds)
	]{\label{fig:ltime}
		\includegraphics[width=0.5\columnwidth]{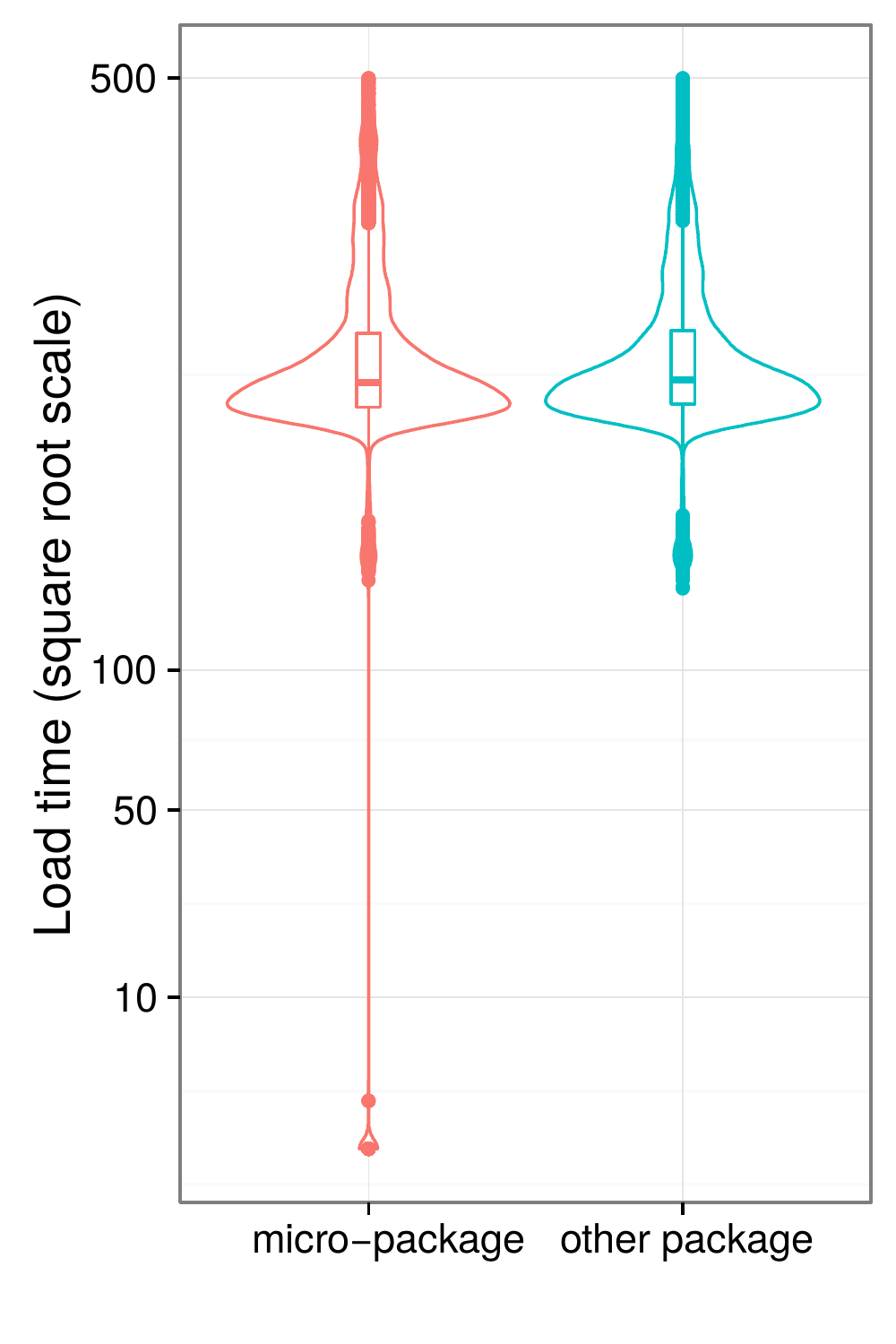}
	}
	\caption{Micro-package developer usage cost metric comparisons against the rest of packages in the \texttt{npm} ecosystem.}
	\label{fig:singleVsMulti}
\end{figure*}

\subsubsection{\underline{Controlled Experiment}}
Algorithm \ref{alg:RQ3scen} details the controlled experiment scenario that simulates the usage of an \texttt{npm} package.
In sequential order, we show how a developer uses the standard protocols to:  (i) fetch an npm package from the NPM ecosystem repositories (\ie~Line 3), (ii) install the package and finally (\ie~Line 4) (iii) load the package into the NodeJs environment using the \texttt{require()} function (\ie~Line 5).
We reset the NodeJs to ensure a controlled environment each time a scenario is run (\ie~Line 2).
To monitor the module loading execution time, we use the \texttt{require-profiler} npm package \cite{NPM:trequire} to measure the execution time for NodeJs the modules loading.

Table \ref{tab:dataExpRQ3} details the technical specifications showing that we were able to successfully install and load 14,126 packages from the 20,000 packages collected.
The controlled experiment was conducted on a \texttt{z620 hp workstation} with eight core-processors running at \texttt{3.60Ghz} and \texttt{32GB} of RAM space, running \texttt{Ubuntu ver.16.04}, and \texttt{NodeJs ver.4.2.6}. 
Using internet speedtests\footnote{speedtest at \url{www.beta.speedtest.net}}, we recorded the internet speed to be at 151.96 MB per second. 
The most common reason for some failed runs was because dependencies were either missing or our NodeJs environment was unsuitable had parameters. 
Overall, the experiment took about 36 hours to execute all 14,126 scenario runs.


\begin{table}
	\begin{center}
		\caption{Developer Usage Cost Statistics of the collected \texttt{npm} (in milliseconds)
		}
		\label{tab:RQ3stats}
		\begin{tabular}{lrrrrrc}
			\hline
			\multirow{1}{*}{} 
			& \multirow{1}{*}{\textbf{Min.}}
			& \multirow{1}{*}{\textbf{Median ($\bar{x})$}}
			& \multirow{1}{*}{\textbf{Mean ($\mu$)}}
			& \multirow{1}{*}{\textbf{Max.}}\\
			Fetch time 	&  1.00 &  109.00 &  113.80 & 623.00 \\ 
			Install time 	&  82.00 &  103.00 &  105.20 & 1,619.00 \\
			Load time & 0.00  &  242.00 &    322.30 &  328,600.00    \\\hline
		\end{tabular}
	\end{center}
\end{table}

\subsubsection{\underline{Findings}}
Table \ref{tab:RQ3stats} shows the summary statistics for the three developer usage cost metrics for \texttt{npm} packages.
We observe from the table that took a median of 109ms to fetch, 105ms to install and 242ms to load a library. 
We find that developers are especially concerned when loading applications with large dependencies (\ie~\textit{`loading around 10.000 files into the NodeJs environment'})\footnote{this discussion shows how developers discuss loading times of JS modules at \url{https://groups.google.com/forum/\#!topic/nodejs/52gksIpgX4Q}}. 
With a manual validation, we find that the slowest loading npm packages had dependencies to large JavaScript libraries. 
For instance, the \texttt{forkeys-benchmark}\footnote{website is at \url{https://github.com/forkeys/forkeys-benchmark}} package took the longest to load into the NodeJs environment. 
After reading documentation, we conclude that the reason for taking the longest time is because its dependency to the JS benchmark library (\texttt{Benchmark.js} \cite{JS:benchmarks}.
Another library that took a significantly long loading time was \texttt{gulp-transformers}\footnote{The package functions to unify gulp, browserify and its transforms once and for all at \url{https://github.com/limdauto/gulp-transformers}}, that has dependencies to the JavaScript \texttt{gulp} \cite{JS:gulp} build tool and the \texttt{browserfly} \cite{JS:browerfly} JavaScript library that lets developers \texttt{require(`modules')} in the browser by bundling up all of dependencies.

\begin{table}[t]
	\begin{center}
		\caption{Statistical Tests of Developer Usage Cost Metrics. $H_0$ states that both library types have significant differences.
		}
		\label{tab:RQ3Comp}
		\begin{tabular}{lrrrrrc}
			\hline
			& \multirow{1}{*}{\textbf{$H_0$}}
			& \multirow{1}{*}{\textbf{p-value}}
			& \multirow{1}{*}{\textbf{d}}\\
			Fetch Time 	&  \colorbox{red!25} {reject} & $<  0.75$ & 0.008 (S) \\ 
			Install Time &  \colorbox{red!25} {reject} & $< 0.49$ & 0.002 (S)\\
			Load Time & \colorbox{red!25} {reject}& $<  0.05$ & 0.012 (S)      \\\hline
		\end{tabular}
	\end{center}
\end{table}

Testing our hypothesis that \textit{micro-packages incur developer usage costs of a micro library are no less than the rest of packages in the \texttt{npm} ecosystem}, we find that micro-packages have no statistically differences in developer usage costs to the rest of the \texttt{npm} packages.
Results in Figure \ref{fig:singleVsMulti} and Table \ref{tab:RQ3Comp} confirm acceptance of the alternative hypothesis that micro-packages share identical developer usage costs as other packages in the ecosystem.

\begin{hassanbox}
	In summary for RQ3, we propose and experiment to investigate developer usage costs for over 20,000 \texttt{npm} packages. To answer RQ3, \textit{our findings indicate that micro-packages have no statistically differences in developer usage costs to the rest of the \texttt{npm} packages.
	}
\end{hassanbox}

\section{Discussion}
\subsection{Implications and Challenges}
\label{sec:dis}
Results of our empirical study contributes to the research performed on software ecosystems \cite{2013CSMRGerman,lungu2008} and their effect on software maintenance.
The issue of micro-packages and how an ecosystem can ripple effects to an application has been studied in related work \cite{Mens:2016}, \cite{Mens:ECOS}.  
However, ongoing research on these self-organizing ecosystems are still unclear on how and why ecosystems form, how they achieve their impact, or how they will continue to sustain themselves.   
Our work is working towards that goal of understanding how to cope with threats to an ecosystem.

The study brings to light many new challenges and avenues for potential future research.
JavaScript best practices\footnote{\url{https://www.w3.org/wiki/JavaScript_best_practices}} recommends to modularize (\ie~limit one function to one task) for better code comprehension and reuse.
However, a byproduct of package modularization is the influx of excessive dependencies in the ecosystem. 
As these ecosystems are self-organizing with no strict social structures, we believe future work is needed to understand how ecosystems to cope with increased dependency complexity over time.
Future work include revisiting the best practices on how to make packages resilient to ecosystem changes whilst maintaining code modularity. 
One potential strategy that is employed in other ecosystems is to merge popular and useful libraries into the core platform package.
For example, the popular \texttt{Joda-time} Java library was migrated into a core part of the Java 8 and onwards JDK (\ie~\texttt{java.time (JSR-310))}.
In addition, we envision that support is needed to raise the awareness of developers to how sensitive their applications are to the ecosystem.

As future work, we would like to extent the notion of packages.
Currently the \texttt{require()} function does not differentiate between internal and external modules. 
Therefore, further work is needed to explore the true notion of the dependency relationships of external modules.
We are however confident that our definition of a micro-package is sufficient to detect this phenomena.

\subsection{Threats to Validity}
\label{sec:threats}


\subsubsection{Construct}
This validity is concerned with threats to our micro-package definitions.
In this work, we simply use the number of functions to reason for a definition of a package, however, in reality the refined definition of a micro-package would be the smallest number of functions in the least amount of code. 
We are confident that our definitions are sufficient to capture excessive dependencies in the ecosystem and discuss how threat is will be treated as future work (\ie~Section \ref{sec:dis}). 

\subsubsection{Internal}
This validity is related to the accuracy of the data collected and tools used in the experiments. One of the major threat is the accuracy of data that is contained in the \texttt{package.json} metafile configuration file for RQ2 and RQ3.
There exists cases when a dependency is not listed in the metafile (\ie~\texttt{require()} call to unlisted dependency module).
We understand this situation can rarely exist, as the vast majority of  quality library would correctly list all dependencies. In future work, we plan to use more direct source-code dynamic analysis to mitigate this threat. Another potential threat can be related to the use of \texttt{esprima} \cite{NPM:esprima} for the generation of code-metrics in RQ1 and \texttt{require-profiler} \cite{NPM:trequire} to measure the developer usage cost metrics in RQ3. To our best knowledge, both these tools are heavily used in the npm ecosystem and are compliant to ECMAScript standards. \texttt{Esprima} runs on many popular web browsers and other ECMAScript platforms such as Rhino, Nashorn, and NodeJs

\subsubsection{External}
The external validity refers to the generalization of our results. We only collected a sample of the true size of JavaScript projects, that are open to the public. Although we believe our large dataset of over 169,000 libraries is a sufficient representation of \texttt{npm} JavaScript current practices, expanding to different ecosystems is a potential avenue for future work.

\section{Related Work}
\label{sec:related}

\subsubsection{Software ecosystems}
Recently, there has been an increase in research that formulate software systems as belonging to an ecosystem.
Lungu best termed ecosystems as a \textit{``collection of software projects which are developed and evolved together in the same environment''} \cite{lungu2008}. There has been other definitions from a business perspective \cite{Jansen:2013, Jansen09ICSE} and also from a technological perspective \cite {Manikas:2013}. In this work, we consider the ecosystem of libraries that belong to the same technological platform, in this case JavaScript. 
Related, Wittern \textit{et al.}~\cite{Wittern:2016} studied the topology of popular \texttt{npm}JavaScript libraries, finding that \texttt{npm} packages are largely dependent on a core set of libraries. 
This is consistent with our result that package reuse is apparent in the ecosystem. 

Related, Mens \textit{et al}~\cite{Mens:ECOS} perform ecological studies of open source software ecosystems related to other software ecosystems.
Haenni \textit{et al.}~\cite{2013:wea:haenni} performed a survey to identify the information that developers lack to make decisions about the selection, adoption and co-evolution of upstream and downstream projects in a software ecosystem. 
Bavota \textit{et al.} \cite{Bavota:2015} has also looked into evolution issues of software ecosystems, particularly in the Apache software ecosystem.
Similar work was performed by German \textit{et al.} \cite{2013CSMRGerman} for the R software ecosystem. 
Robbes \cite{Robbes:2012} studied the API deprecations gravitating around the Squeak and Pharo software ecosystems. 
They found that dependency changes such as API deprecations can indeed have a very large impact on the ecosystem. 
This further illustrates how much impact an disruption to the software ecosystem can be for both library users and developers.

\subsubsection{Library Reuse and Maintenance}
Serebrenik \cite{Serebrenik:2015} and Mens \cite{Mens:2016} discussed the many challenges that face software ecosystem, especially in terms on evolution updates and stability of dependencies. In summary, most work on library reuse has been on software library breakages, library updating and migration to stable versions. For instance, Bogart \textit{et al.} \cite{Bogart:SCGSE15} studied how developers reacted to stability of dependencies in their ecosystem. 
Other existing work are related to the traditional java and C programming language. For instance, Teyton \textit{et al.} \cite{Teyton:2014} studied library migration for Java systems.
In this work they studied library migrations of java open source libraries from a set of client with a focus on library migration patterns. Another work was by Xia \textit{et al.} \cite{Xia2013}, that studied the reuse of out-dated project written in the c-based programming language. 
Recently, there has been large-scale empirical studies conducted on library migrations and evolution. 
Raemaekers \textit{et al.} \cite{RaemaekersICSM} performed several empirical studies on the Maven repositories about the relation between usage popularity and system properties such as size, stability and encapsulation.
They also studied the relationship between semantic versioning and breakages \cite{Raemaekers2014}.
Other related empirical studies were conducted by Jezek \textit{et al.} \cite{Jezek2015} and Joel \textit{et al.} \cite{Cox:2015}. They studied in-depth how libraries that reside in the Maven Central super repository evolve. In this work, we are concerned with the potential possibilities of library maintenance involved for micro-packages. So far this issue has only been reported by the JavaScript \texttt{npm} community, however, smaller designs may not necessarily mean less library support.
 
\section{Conclusion}
\label{sec:conclusion}
In this paper, we explore the role of a micro-package in the \texttt{npm} JavaScript ecosystem. 
Results find that micro-packages form a significant portion of the \texttt{npm} ecosystem. Apart from the ease of readability and comprehension, we show that some micro-packages having long dependency chains incur just as much usage costs as the rest of the packages. 

We envision that this work motivates the need to further investigate the notion of micro-packages and how to increase package resilience to changes in the ecosystem. 
The study concludes that it is important for developer to be aware of ecosystem changes that might affect their applications.

\bibliographystyle{abbrv}
\bibliography{sigproc}

\end{document}

%% file: macros.tex
\newenvironment{hassanbox}%
{\begin{center}\vspace{0mm}\noindent\begin{Sbox}\begin{minipage}{0.95\columnwidth}}%
{\end{minipage}\end{Sbox}\fbox{\TheSbox}\end{center}\vspace{0mm}}